\documentclass[lettersize,journal]{IEEEtran}
\usepackage{amsmath,amsfonts}
\usepackage{algorithmic}
\usepackage{algorithm}
\usepackage{array}
\usepackage[caption=false,font=normalsize,labelfont=sf,textfont=sf]{subfig}
\usepackage{textcomp}
\usepackage{stfloats}
\usepackage{url}
\usepackage{amssymb}   % for \checkmark
\usepackage{tabularx}
\usepackage{array}
\usepackage{booktabs}
\usepackage{caption}
\usepackage{subfig}
\usepackage{verbatim}
\usepackage{graphicx}
\usepackage{multirow}
\usepackage{cite}
\usepackage{multirow}%
\usepackage{amsmath,amssymb,amsfonts}%
\usepackage{xcolor}%
\usepackage{booktabs}%
\usepackage{subfig}
\usepackage{soul}
\begin{document}

\title{ATRACT: A Trustworthy Robotic
Autonomous system to support Casualty Triage}

\author{
Tasweer~Ahmad\textsuperscript{\textdagger},
Rafael~Pina\textsuperscript{\textdagger},
Sandip~Pradhan\textsuperscript{\textdagger},
Arindam~Sikdar,
Mindula~Illeperuma,
Khizer~Saeed,
Peter~Lee,
Varuna~De~Silva,
Ardhendu~Behera\textsuperscript{*}%
\noindent
\thanks{\textsuperscript{\textdagger}Tasweer Ahmad, Rafael Pina, Sandip Pradhan equally contributed.}%
\thanks{\textsuperscript{*}Corresponding author: Ardhendu Behera (e-mail: beheraa@edgehill.ac.uk).}%
\thanks{Tasweer Ahmad, Sandip Pradhan, Arindam Sikdar, Ardhendu Behera are with the Department of Computer Science, Edge Hill University, Ormskirk, U.K. (e-mail: Tasweer.Ahmad@edgehill.ac.uk; Sandip.Pradhan@edgehill.ac.uk; Arindam.Sikdar@edgehill.ac.uk; beheraa@edgehill.ac.uk).}%
\thanks{Rafael Pina, Mindula Illeperuma, Varuna De Silva are with the Institute for Digital Technologies, Loughborough University London, 3 Lesney Avenue, London E20 3BS, U.K. (e-mail: r.m.pina@lboro.ac.uk; k.m.illeperuma@lboro.ac.uk; v.d.de-silva@lboro.ac.uk).}%
\thanks{Khizer Saeed is with the School of Architecture, Technology and Engineering, University of Brighton, U.K. (e-mail: K.Saeed@brighton.ac.uk).}%
\thanks{Peter Lee is with the School of Criminology and Criminal Justice, University of Portsmouth, Portsmouth, U.K. (e-mail: peter.lee@port.ac.uk).}%
}

\markboth{Journal of \LaTeX\ Class Files,~Vol.~14, No.~8, August~2021}%
{Shell \MakeLowercase{\textit{et al.}}: A Sample Article Using IEEEtran.cls for IEEE Journals}


\maketitle

\begin{abstract}
At a time when drones are increasingly associated with hostile operations, we re-purpose them for humanitarian and life-saving applications. However, adapting search and rescue drones for battlefield triage remains extremely challenging; the technology must perform reliably to support frontline medics who are forced to operate under extreme uncertainty, restricted access, and significant personal risk. Due to growing vulnerabilities of casualty evacuation in conflicting zones, this paper presents \textit{ATRACT} (A Trustworthy Robotic Autonomous system to support Casualty Triage), a novel human-in-the-loop decision-support system to enable early battlefield triage during the critical post-trauma period. ATRACT integrates drone-captured video with wearable sensor input for multi-modal learning to support casualty-state assessment, thereby addressing the limitations of existing systems. 
Drone video captures fine-grained behavioural cues, such as pose, posture, while body-worn sensors provide complementary physiological signals, including heart rate, breathing rate, and movement. By combining two modalities, ATRACT provides evidence to support the early judgement of medics when direct access to the casualty is delayed, risky, or restricted. To mitigate the data realism gap pertaining to injured actions, a conditional variational autoencoder is devised for data augmentation. 
Experimental results on our drone-captured dataset show that proposed pipeline achieves 85.7\% accuracy
for action classification; while our lightweight CNN visual
encoder remains competitive with stronger pre-trained video
backbones. 
Overall, the results support ATRACT as a practically meaningful step towards remote triage in contested environments, where multi-modal sensing, human oversight and trustworthy decision support can improve casualty prioritisation, and lessen the exposure of frontline medics.
\end{abstract}

\noindent

\section{Introduction}
\IEEEPARstart{T}{he} Vietnam War established the principle of ``Golden Hour'', revealing that rapid evacuation and early intervention can significantly improve survival, a \emph{doctrine} that has since shaped UK, US, and NATO casualty care. However, recent conflicts, e.g., Ukraine, have exposed the increasing vulnerability of helicopter-based casualty evacuation (CASEVAC) to low-cost, highly vulnerable air-defence threats. This forces the frontline medics to operate under extreme risk. %while frontline medics have to undergo for triage in extreme risk, uncertainty, and operational pressure. 
The above challenge is particularly critical within the ``platinum ten minutes'' immediately after trauma, when timely situational awareness can significantly increase the chances of survival. 

In this context, timely casualty assessment in the battlefield remains a challenging problem, particularly when access to injured persons is restricted and triage decisions must be promptly made under limited medical information. In such settings, even a short delay in health state assessment (e.g., injured, un-responsiveness) can badly affect the triage prioritisation, planning, and ultimate survival. Yet manual triage remains cognitively demanding, resource intensive, and potentially dangerous for responders operating in contested areas. More broadly, recent work in conflict scenarios has highlighted the growing importance of technology-driven remote medical support, with \emph{telemedicine} emerging as a promising mechanism to extend the clinical expertise under disrupted healthcare infrastructure and limited on-site resources \cite{habers2025telemedicine}. These challenges motivate the development of intelligent remote-sensing systems that can observe, interpret, and summarise casualty condition without requiring immediate physical contact. In this context, the \emph{ATRACT} investigates a human-in-the-loop triage system in which drone-driven active-sensing with wearable sensor inputs are integrated for multi-modal learning to support the rapid and reliable assessment of injured and uninjured soldiers in the battlefield scenarios.\\
\textbf{RQ:} Accordingly, this research \emph{questions} whether an integrated triage system, combining active drone sensing, multi-modal learning, and human-in-the-loop oversight, can support a trustworthy, and operationally meaningful remote casualty assessment in contested environments?

In recent years, there has been a shift towards using technology-assisted emergency assessment~\cite{alvarez2022victim, mosch2024automated}. UAVs and drones have increasingly been explored for victim detection~\cite{andriluka2010vision, al2019life}, situational awareness, and triage support due to rapid coverage to wide areas, access hazardous locations, and acquire aerial views without exposing rescuers to the same level of risk. More recent work has moved beyond simple victim localisation towards contactless assessment of clinically relevant cues, (e.g., body pose, consciousness, breathing rate) and semi-automatic triage from the aerial images, \cite{al2019life, queiros2021consciousness}. These advances anticipate that remote casualty assessment is no longer merely conceptual; it is becoming technically feasible. However, existing systems remain largely \emph{fragmented}; mostly focus on either victim localization, life-sign detection, or semi-automatic triage in civilian mass-casualty scenarios~\cite{mosch2024automated}. Such systems are unable to fully address the battlefield-specific problems in which injury, immobility, and physiological state pose a real challenge for remote triage.

\begin{figure*}
    \centering
    \includegraphics[width=0.95\textwidth]{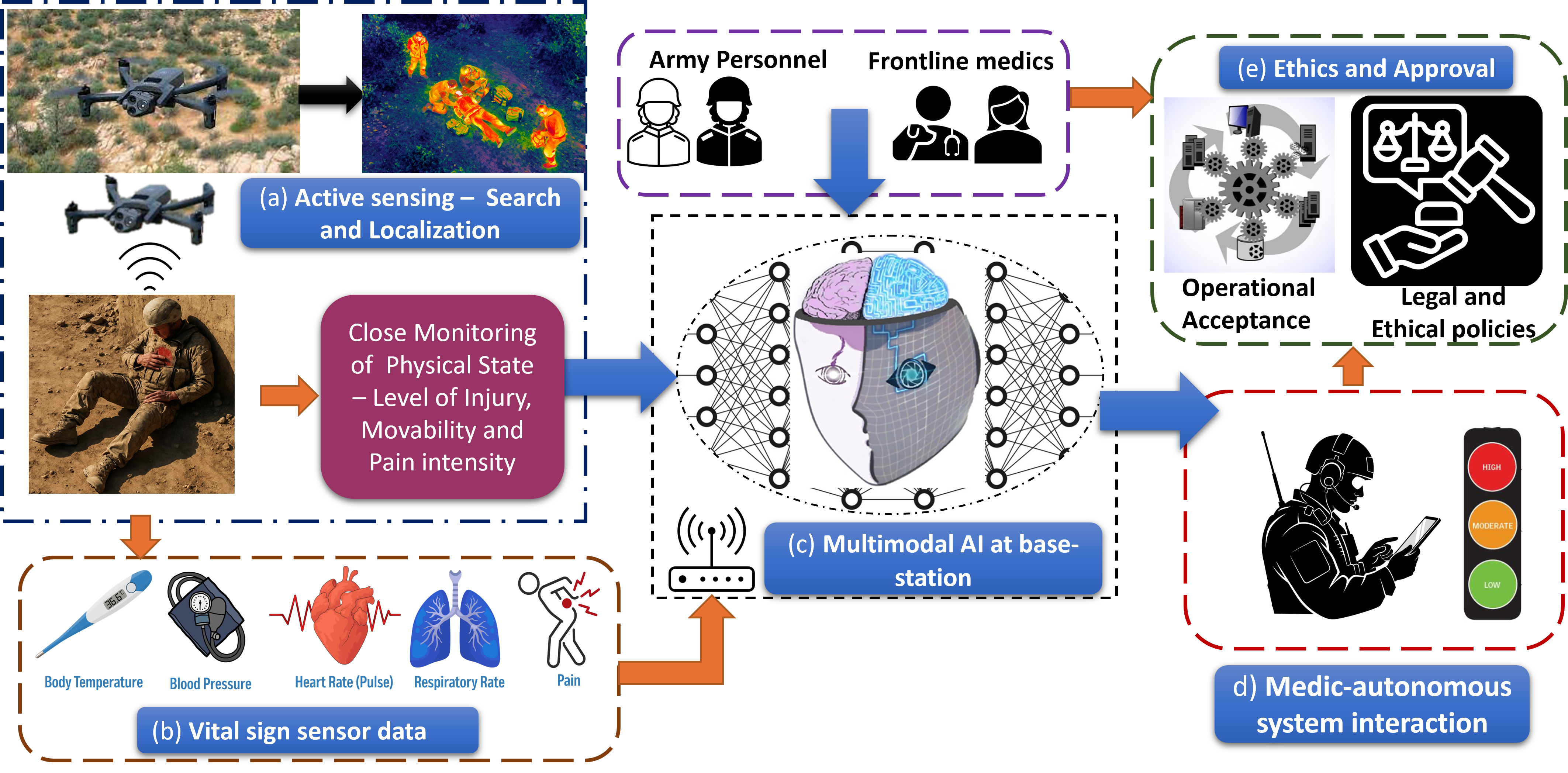}
    \caption{Overview of the proposed ATRACT framework for human-in-the-loop battlefield triage. The system uses \textbf{(a)} \emph{active sensing} drone for search and localisation, \textbf{(b)} \emph{vital signs monitoring} using wearable sensors, and \textbf{(c)} data fusion for \emph{multi-modal learning}. \textbf{(d)} \emph{Inference} categorises triage output as low, mild and severe injury. \textbf{(e)} The final decision is constrained by \emph{human-in-the-loop} oversight. Overall, the framework illustrates a novel end-to-end pipeline for remote and scalable casualty assessment in a contested environment.}
    \label{fig:ATRACTconceptdiagram}
\end{figure*}

\noindent
\textbf{Motivation:} The central premise of the ATRACT system is an integrated casualty assessment using different modalities for trustworthy AI-assisted decision support. Drone videos contribute to an important data modality and reveal motion dynamics in the scene, (e.g., subject movement, body posture, pose etc.); where prior work evidenced that aerial images can support victim detection and consciousness assessment~\cite{andriluka2010vision,queiros2021consciousness,mosch2024automated}. However, visual evidence alone is brittle under aerial viewpoints, occlusion, clutter, motion blur, scale and viewpoint variation~\cite{andriluka2010vision}. In contrast, body-worn sensors provide complementary information (heart rate, breathing rate, body posture, movement), as physiological or behavioural indicators are less evident in videos yet highly informative for assessing the physical condition. This complementarity is also consistent with the recent advances in contactless life-sign sensing, where cardiopulmonary motion and heart rate have been shown to provide clinically relevant information for remote assessment in disaster and triage scenarios \cite{al2019life}. ATRACT, therefore, adopts a novel multi-modal approach in which \emph{fine-grained behavioural clues} from drone-captured videos are fused with \emph{vital signs} from wearable sensors, enabling a robust and operationally meaningful remote triage support. % than either modality can provide individually. 
This design choice reflects a principled response to the uncertainty inherent in real triage, where ambiguous visual behaviour may be clarified by vital signs, and noisy physiological signals may be contextualised into motion patterns.

% The ATRACT system faces a unique challenge of scarce realistic data for the injured-person. In the first phase, we collected data during a field exercise, where the participant performed six different actions (e.g., arm injury, head injury, limping, etc.), while wearing the body sensors. This sensor data is augmented using a \emph{Conditional Variational Vital Auto Encoder (CVVitAE)} \cite{cvvitae_2025}, which learns from the ICU-trained clinical data and augments the vital signs to healthy participants in drone videos. This fills the data realism gap, since gathering injured-soldiers physiological data in a controlled experiments is inherently difficult, ethically constrained, and operationally unrealistic. %Thus, the project addresses not only the sensing and fusion problem, but also the data realism problem that often limits the progress in this domain.

Within the ATRACT framework, raw multi-modal video and sensor streams are progressively transformed into useful evidence for battlefield triage, as detailed in Fig.~\ref{fig:ATRACTconceptdiagram}. At high level of conceptualisation, the drone performs active sensing and localisation, followed by extracting person-centric trajectories in videos using detection and tracking. In parallel, wearable sensors record physiological and behavioural measurements and transmit to the base station for monitoring and late fusion. The visual stream is encoded to capture spatio-temporal motion cues, while the sensor stream is modelled to capture temporal physiological dynamics; such modality-specific representations are concatenated for person-state recognition. \\
\noindent
\textbf{Contribution:} Overall, ATRACT brings a practical deployable AI system for remote casualty assessment. Its contribution lies not only in establishing an end-to-end multi-modal sensing-and-learning pipeline, but also framing ATRACT as a triage support system that integrates active sensing, human oversight, and remote assessment.
% Its contribution lies mainly in establishing an end-to-end framework that combines active drone sensing with physiological inputs for multi-modal learning and decision support within autonomous triage. 
By jointly reasoning over motion, posture, and vital-sign behaviour, the project aims to judge the casualty health state and provide a stronger foundation for future decision-support systems in challenging environments. \\
Next, related work is covered in Section~\ref{related work} and ATRACT system components in Section~\ref{framework}. The results are discussed in Section~\ref{Results_Sec2}, with explainability and complexity in Sections~\ref{label:explainability} and~\ref{complexity} respectively. Ethical concerns are set out in Section~\ref{sec:ethics}, while this research is concluded in Section~\ref{conclude}.

% The following methodology section details the data acquisition pipeline, the proposed generative model for physiological signal adaptation, the visual encoding and tracking components, and the late-fusion multi-modal learning framework used to realise this vision. The manuscript is organized to summaraize the related work in Section~\ref{related work}, preliminaris of the project in Section~\ref{preliminaries}. The detailed framework of our ATRACT syetem is explained in Section~\ref{framework}. We discuss the experimental setup in Section~\ref{experimental setup}, while the results are discussed in Section~\ref{Results_Sec2}. We finally draw conclusion in Section~\ref{conclude}.

% \vspace{-0.1cm}

\section{Related Work}\label{related work}
\subsection{Pre-hospital Triage Systems}
\noindent
Triage is a cornerstone of both pre-hospital emergency care and mass casualty incident management. Conventional scoring instruments such as Revised Trauma Score~\cite{Champion1989}, or the Injury Severity Score~\cite{Baker1974} encode expert clinical knowledge into rule-based frameworks. While still useful, their reliance on manual assessment makes them unsuitable for critical environments~\cite{Lerner2011}. On the other hand, early work~\cite{christie2018machine} demonstrated that machine learning classifiers trained on trauma data could predict mortality across diverse settings with accuracy exceeding conventional scoring systems, highlighting the potential of data-driven methods to augment clinical judgement. In the context of battlefield triage, the TCCC (Tactical Combat Casualty Care) framework defines clinical priorities of haemorrhage control, airway management, and rapid evacuation~\cite{Butler2000}; an automated system capable of supporting priorities under austere conditions is an active area of research~\cite{Convertino2020}. Wearable sensor approaches with algorithmic decision support have demonstrated promising results for remote patient physiological monitoring~\cite{Pantelopoulos2010}. 
Drones and UAVs have supported pre-hospital outdoor triage in terms of early situational awareness and their rapid search-and-rescue ability in inaccessible areas provide a decisive operational advantage~\cite{Erdelj2017}. 
Likewise, there is a growing usage of small drones for medical delivery~\cite{Haidari2016} to support remote triage~\cite{alvarez2021development}. The distinctive battlefield environment imposes constraints that existing pre-hospital triage is not designed to meet. Therefore, ATRACT builds upon foundations to provide a comprehensive triage in a very challenging battlefield environment.

\vspace{-0.2cm}
\subsection{Action Recognition in Drone Video}

Action recognition in drone captured videos remains difficult due to changing viewpoint and person-centric aspect-ratio, motion blur and occlusion due to obstructed views that weaken the visual cues needed for reliable recognition. Okutama-Action~\cite{barekatain2017okutama} first defined this problem by introducing a UAV benchmark with concurrent dynamic action transitioning, underpinning the need for action recognition in drone videos. The research progresses by~\cite{perera2019drone} introducing an outdoor high-resolution drone video dataset to better capture body details for action recognition. The research findings established that action analysis in drone videos is challenging due to perspective distortion, occlusion, and camera motion. DroneCaps~\cite{algamdi2020dronecaps} introduced capsule networks and binary volume features to better handle pose and viewpoint variation, small human targets, and multi-label actions in drone videos. UAV-Human~\cite{li2021uav} broadened the field to a large multi-task UAV dataset, and highlighted \emph{fisheye distortion} as a distinct challenge in aerial video. Unlike previous work focused on RGB clips, it paired the benchmark with guided Transformer-I3D design to better handle distorted fisheye UAV videos. PMI Sampler~\cite{xian2024pmi} showed that adaptive selection of motion-rich frames can improve recognition when humans occupy only a small number of frames. While Drone-HAT~\cite{khan2024drone} combined detection, tracking, and a hybrid attention transformer with multi-scale and multi-granularity fusion to better recognize multi-label actions in drone surveillance videos. 
Aerial human pose estimation is closely related to action recognition in drone videos because top-down views, small subject size, and self-occlusion weaken the body cues needed for reliable action understanding. In this line, AirPose~\cite{saini2022airpose} addressed multi-view 3D pose estimation from uncalibrated UAV cameras, Active Human Pose Estimation~\cite{chen2024active} addressed best-view planning for better pose recovery, and FlyPose~\cite{farooq2026flypose} addressed lightweight real-time top-down pose estimation in aerial videos.

Collectively, the above studies establish that reliable action recognition in drone videos depends on robust modelling of viewpoint variation, occlusion, small human subjects, and body pose perspective, which motivates ATRACT to use drone video as a complementary modality to examine the spatio-temporal backbones for action recognition. % -----Within ATRACT, the visual input provides complementary cues (e.g., body posture, movement etc.) to recognise different human actions (injured/un-injured) in drone videos. Unlike image, video-based action recognition depends on modeling both the spatial appearance and temporal evolution, since many injury states unfolds over the time. Earlier action-recognition pipelines typically relied on handcrafted motion descriptors (e.g., optical flow, trajectory features~\cite{wang2013action, mahbub2011optical, bobick2002recognition}), whereas contemporary deep learning methods learn end-to-end hierarchical spatio-temporal representations directly from video data~\cite{feichtenhofer2020x3d, arnab2021vivit, liu2022video}. To effectively model combat related actions in our dataset, it is imperative to capture both spatial appearance and temporal motion over the video clips. 
In this line, \emph{3D convolutional networks} jointly learn the spatio-temporal features directly from the video volumes. Earlier models, C3D~\cite{tran2015learning} showed that homogeneous stacks of 3D kernels can learn generic spatio-temporal representations for video classification; while I3D later demonstrated that 2D image backbones can be \emph{inflated} into 3D architectures to further improve the large-scale video action recognition~\cite{carreira2017quo}. %Following work on 3D-CNN models focused on designing spatio-temporal architectures that better balance the benefit of temporal modeling with computational efficiency.
Tran \emph{et al.}~\cite{tran2018closer} systematically examined and introduced powerful variants using residual video networks. In particular, R3D applies full 3D convolutions throughout the network; MC3 reduces the computational cost by combining 3D convolutions in earlier layers with 2D convolutions in later layers; while R(2+1)D factorises a 3D convolution into separate spatial and temporal operations, which helps decoupled spatial and temporal learning and often improves optimisation while keeping the model's complexity comparable. Using efficient design, S3D separated spatio convolutions from temporal to balance speed and performance. % [More recent work has focused on architectural choices that balance temporal modelling capacity and computational efficiency. Tran \emph{et al.}~\cite{tran2018closer} systematically studied spatio-temporal convolutions within residual networks and introduced representative variants that are now widely used as standard video backbones: \textbf{R3D}, which applies full 3D convolutions throughout the network; \textbf{MC3}, which mixes 3D convolutions in earlier stages with cheaper 2D convolutions in later stages to reduce computational cost while retaining motion sensitivity; and \textbf{R(2+1)D}, which factorises a 3D convolution into a 2D spatial convolution followed by a 1D temporal convolution. This factorisation decouples spatial and temporal learning and increases non-linearity, which often improves optimisation and temporal representation learning at comparable complexity. In parallel, efficient architectures such as \textbf{S3D} bring favourable speed--accuracy trade-offs by employing separable spatio-temporal operators.]
Above-mentioned spatio-temporal backbones serve as visual encoder for feature extraction and are \emph{pre-trained} on large-scale datasets (e.g., Kinetics-400) and thus helpful for ATRACT, with limited drone videos for model training. %These pre-trained backbones (S3D, R3D, MC3, and R(2+1)D using Kinetics-400 dataset) served as visual encoder for feature extraction.

% [In practice, these video backbones are commonly initialised from \emph{pre-trained weights} learned on large-scale action datasets, since training 3D CNNs from scratch is both data-hungry and computationally expensive. TorchVision provides Kinetics-400 pre-trained weights for Video-ResNet variants (e.g., R3D-18, MC3-18, R(2+1)D-18) and S3D, and reports their accuracy and computational characteristics (parameters and GFLOPs). Above considerations is particularly relevant for our ATRACT, where drone-captured datasets provide a limited number of participants and requires substantial preprocessing, annotation, and curation prior to model training. We therefore adopt established Kinetics-pretrained video backbones (S3D, R3D, MC3, and R(2+1)D) as the visual encoder, and perform task-specific adaptation before late fusion with physiological sensor signals for autonomous battlefield triage.]

% \vspace{-0.25cm}

\subsection{Multi-modal learning}
In multi-modal learning, different input modalities are systematically integrated to complement each other for improving the robustness of an AI system. Early fusion in safety critical camera--radar applications~\cite{nobis2019deep}, showed that multi-level feature fusion with careful spatial--temporal alignment, can yield more reliable perception under noisy and partial modality degradation. In medical applications, late-fusion combines wearable physiological signals (e.g., ECG, respiration, IMU) with speech cues, and reports that decision-level fusion is strongly influenced by cross-modal synchronisation while remain practical for real-time deployment~\cite{xefteris2023multimodal}. A recent multi-modal approach for edge-devices~\cite{yang2025edge} further emphasises lightweight, synchronised fusion for real-time performance under computational constraints. 
Alongside, graph-based Fusion-GCN has been explored for action recognition~\cite{duhme2021fusion}, which extends skeleton-based GCNs by injecting additional modalities (e.g., IMU and RGB-derived features) into skeleton graph, either as extra node attributes (channel fusion) or as new sensor nodes connected to body joints (spatial fusion). %Their results suggest that graph-level fusion can improve robustness, while also highlighting that fusion design is critical; adding incompatible modalities or fusing along an unsuitable dimension can degrade performance.
Closely, \emph{sensor-to-vision} action recognition studied knowledge distillation~\cite{liu2021semantics}, where different wearable streams (e.g., acceleration, gyroscope, orientation) act as teaching modalities to improve RGB-video stream (a student) through distilling semantic attention into the video network. This sensor-guided knowledge transfer is also aligned with ATRACT's multi-modal concept to reinforce the cross-modal interaction between wearable sensor and video inputs.
\section{The ATRACT System} \label{framework}

\begin{figure*}[!t]
    \centering
    \includegraphics[width=0.8\textwidth]{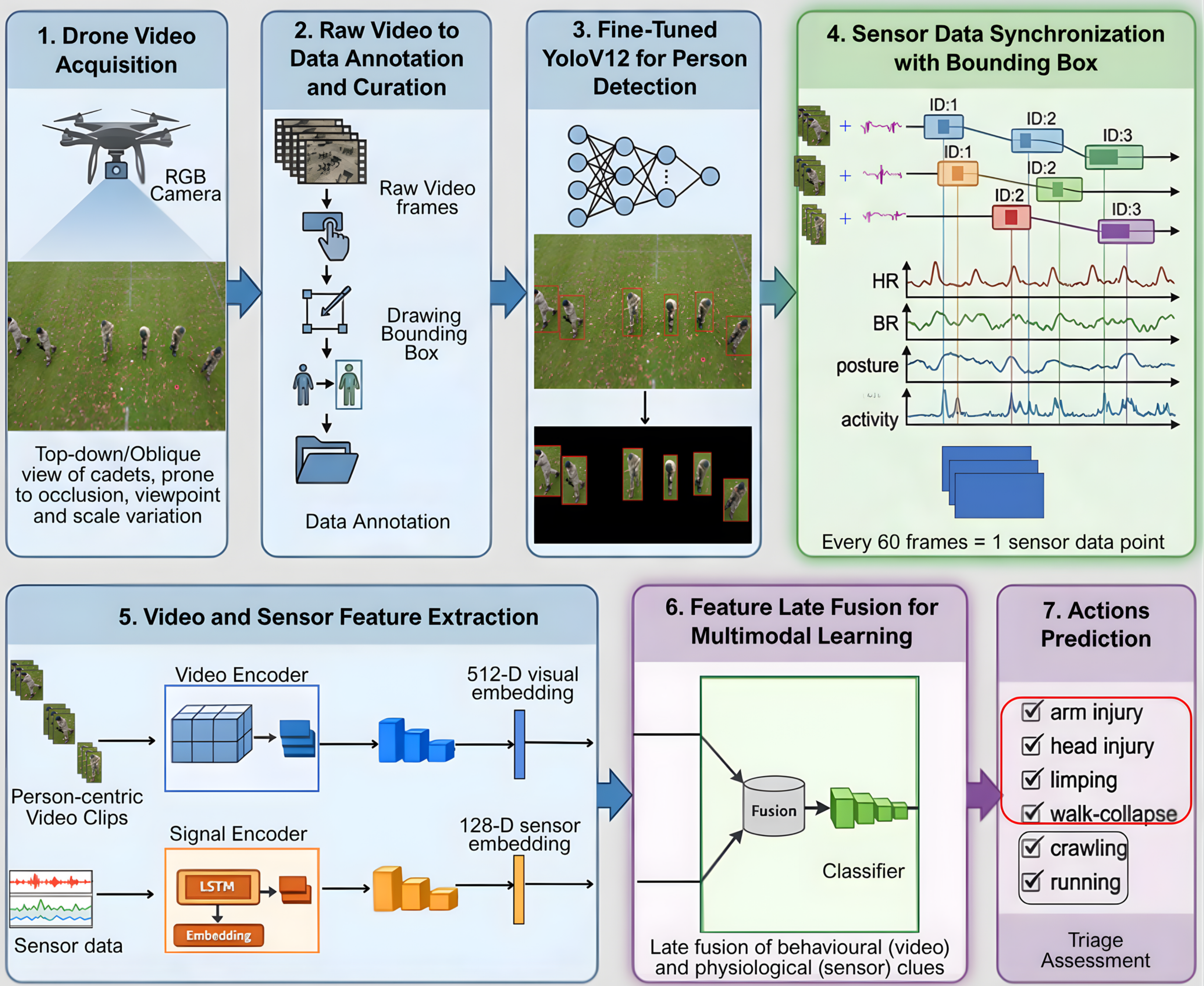}
    \caption{A step-by-step pipeline of our human-in-the-loop ATRACT system. The architecture includes \textbf{1)} drone data acquisition, \textbf{2)} annotation, \textbf{3)} fine-tuning and \textbf{4)} synchronization. Then, there is \textbf{5)} video and sensor feature extraction, \textbf{6)} late fusion for multi-modal learning, classification and \textbf{7)} prediction of injured (arm, head, limping, walk-collapse) and un-injured actions (crawling, running).}
    \label{fig:final flow}
\end{figure*}

Our framework integrates sensor and drone video inputs for human-in-the-loop triage support system, shown in Fig~\ref{fig:final flow}. The body-worn sensor data is pre-processed and synchronised for feature encoding, while features are extracted from videos; fusing both modalities to recognise casualty health-state. %ATRACT workflow is shown in Fig~\ref{fig:final flow}.

% Our proposed ATRACT multi-modal system pipeline involves sensor and video input modalities, where body worn sensors data is pre-processed, synchronized and passed to LSTM for feature encoding, whilst video frames are processed through an encoder for features extraction; both sensor and video features are concatenated for action classification, shown in Fig~\ref{fig:final flow}.

\begin{figure}[!t]
    \centering
    \includegraphics[width=\columnwidth]{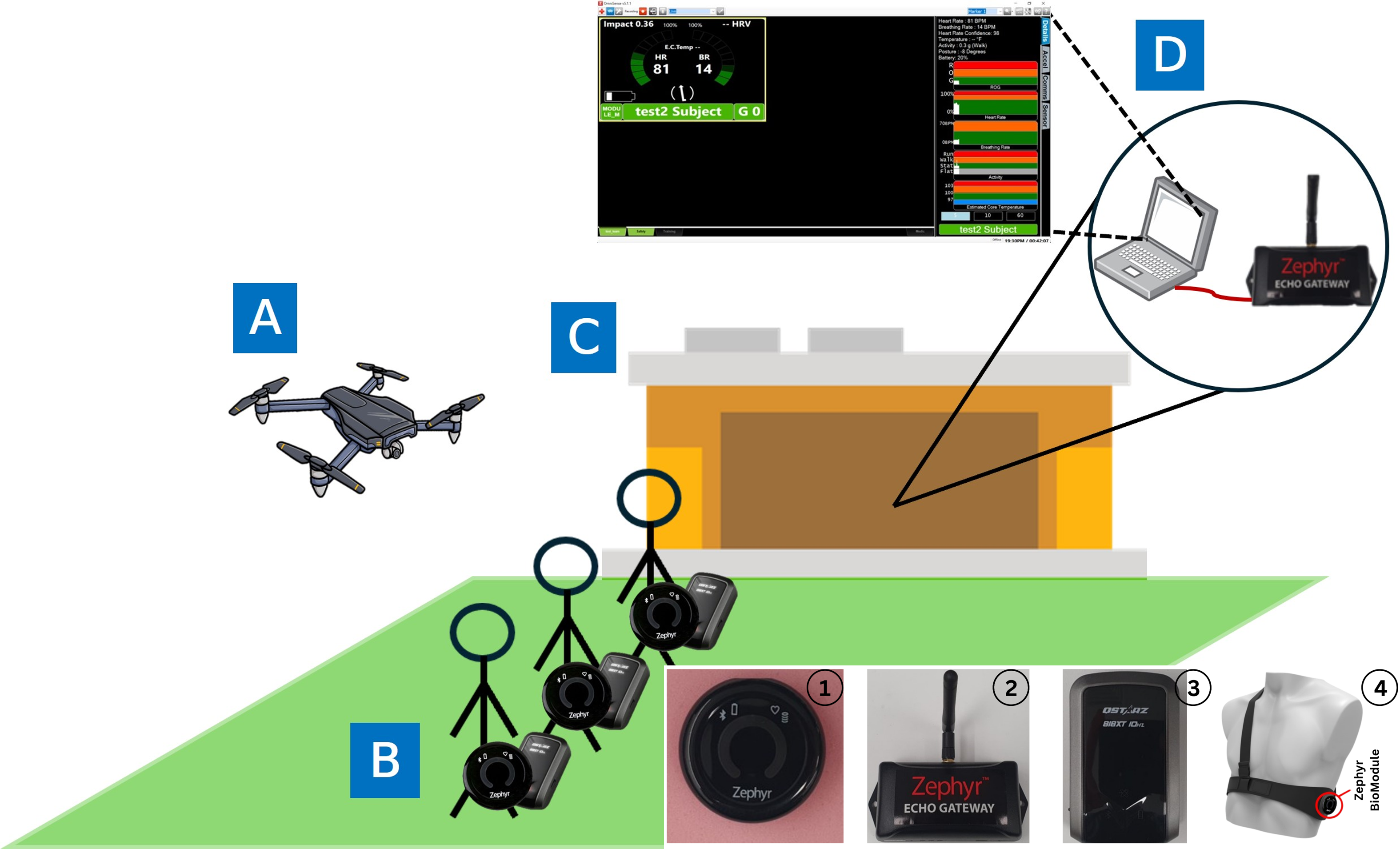}
    \caption{Data collection setup: \textbf{(A)} DJI Mini 2 drone, \textbf{(B)} participants wearing BioModules, \textbf{(C)} base station, \textbf{(D)} data logging using Omnisense~\cite{zephyrtechnology_omnisense}. 
    Sensors used: \textbf{1)} Zephyr BioModule, \textbf{2)} Echo Gateway Transceiver, \textbf{3)} GPS (Starz 818XT), and \textbf{4)} BioModule on chest strap.}
    \label{fig:data_collection_and_sensors}
\end{figure}

\begin{figure*}[t]
    \centering
    \subfloat[Arm Injury\label{fig:scene_a}]{
        \includegraphics[width=0.152\textwidth]{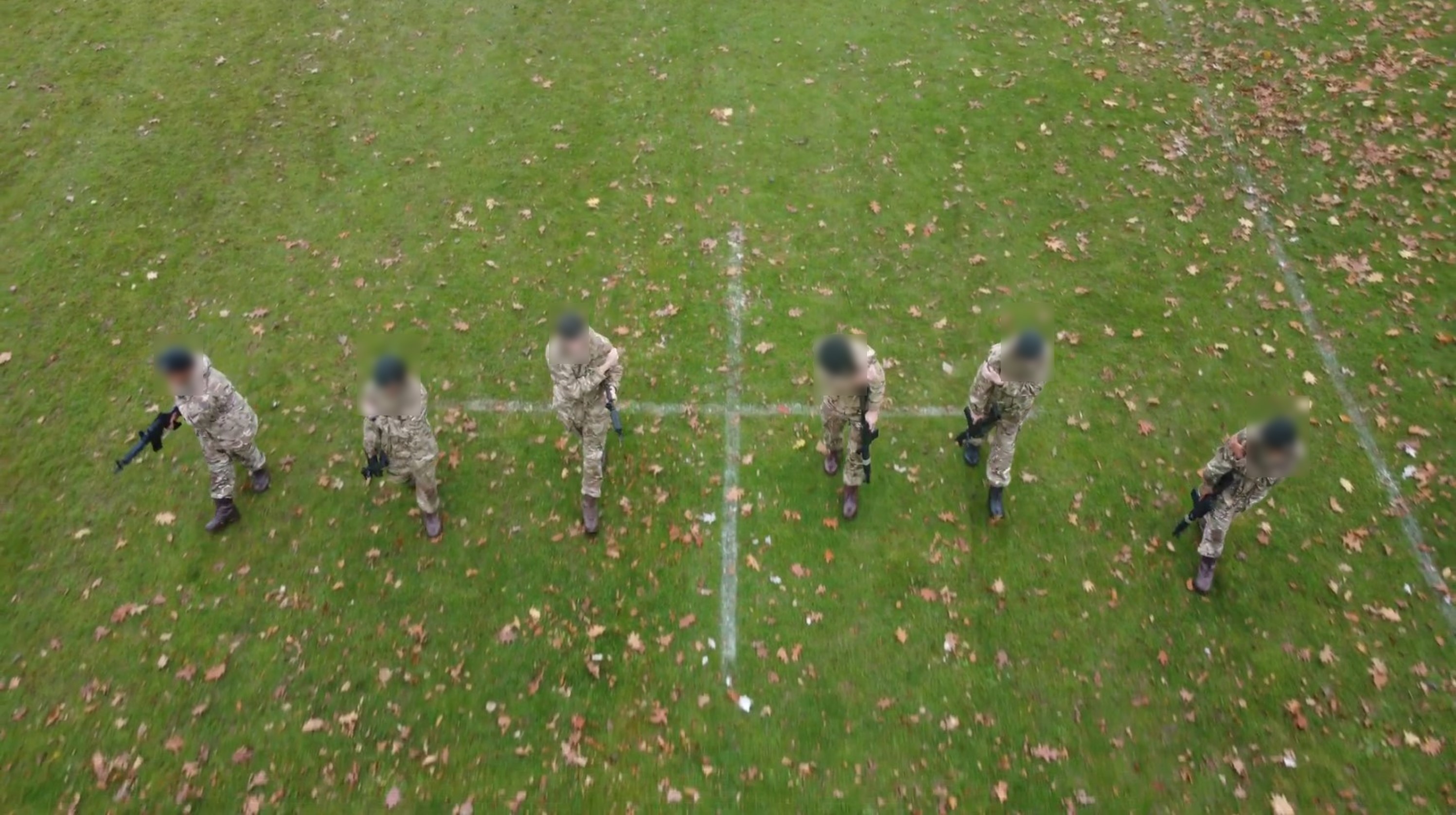}
    }\hfill
    \subfloat[Head Injury\label{fig:scene_c}]{
        \includegraphics[width=0.152\textwidth]{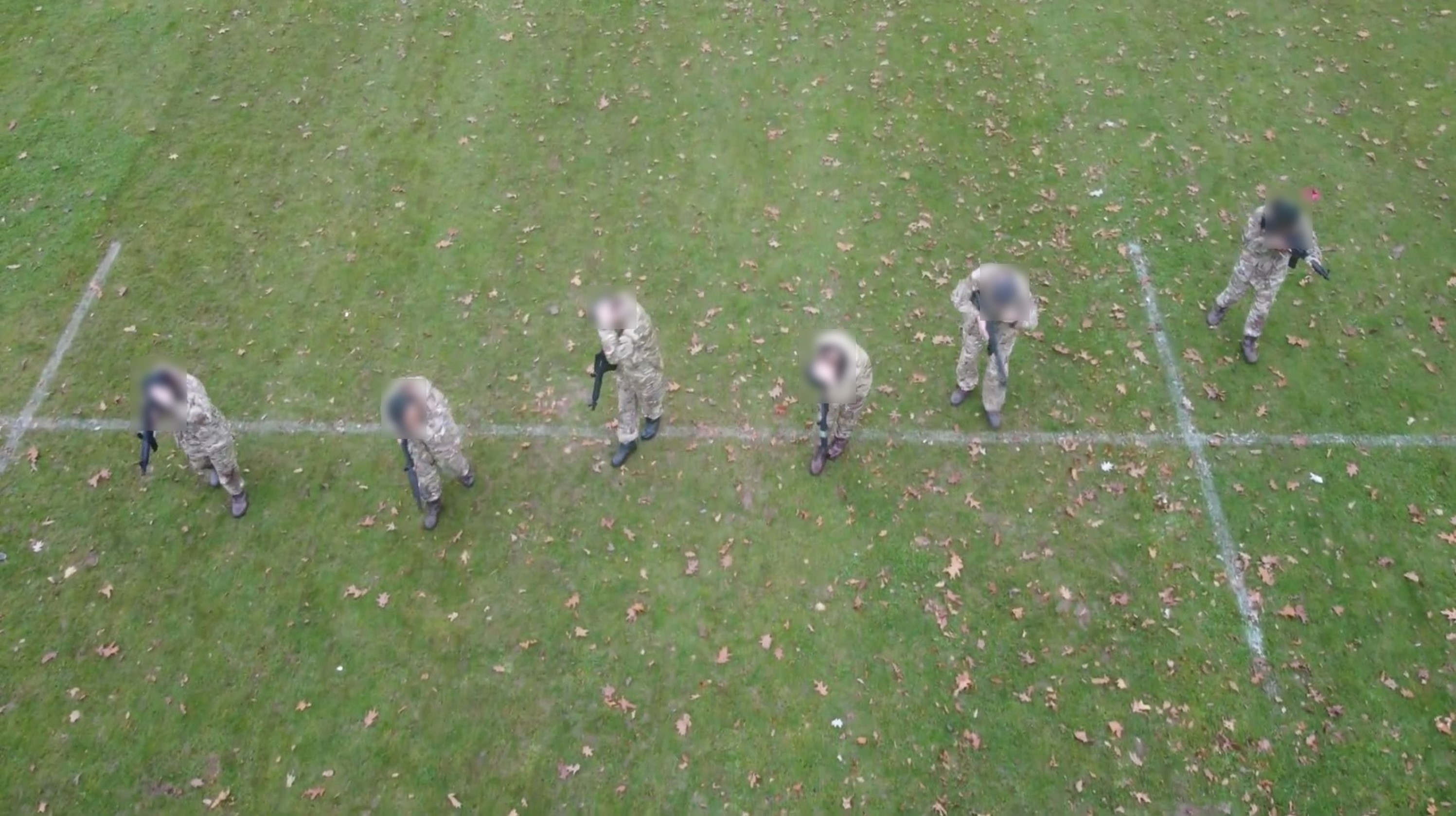}
    }\hfill
    \subfloat[Limping\label{fig:scene_d}]{
        \includegraphics[width=0.152\textwidth]{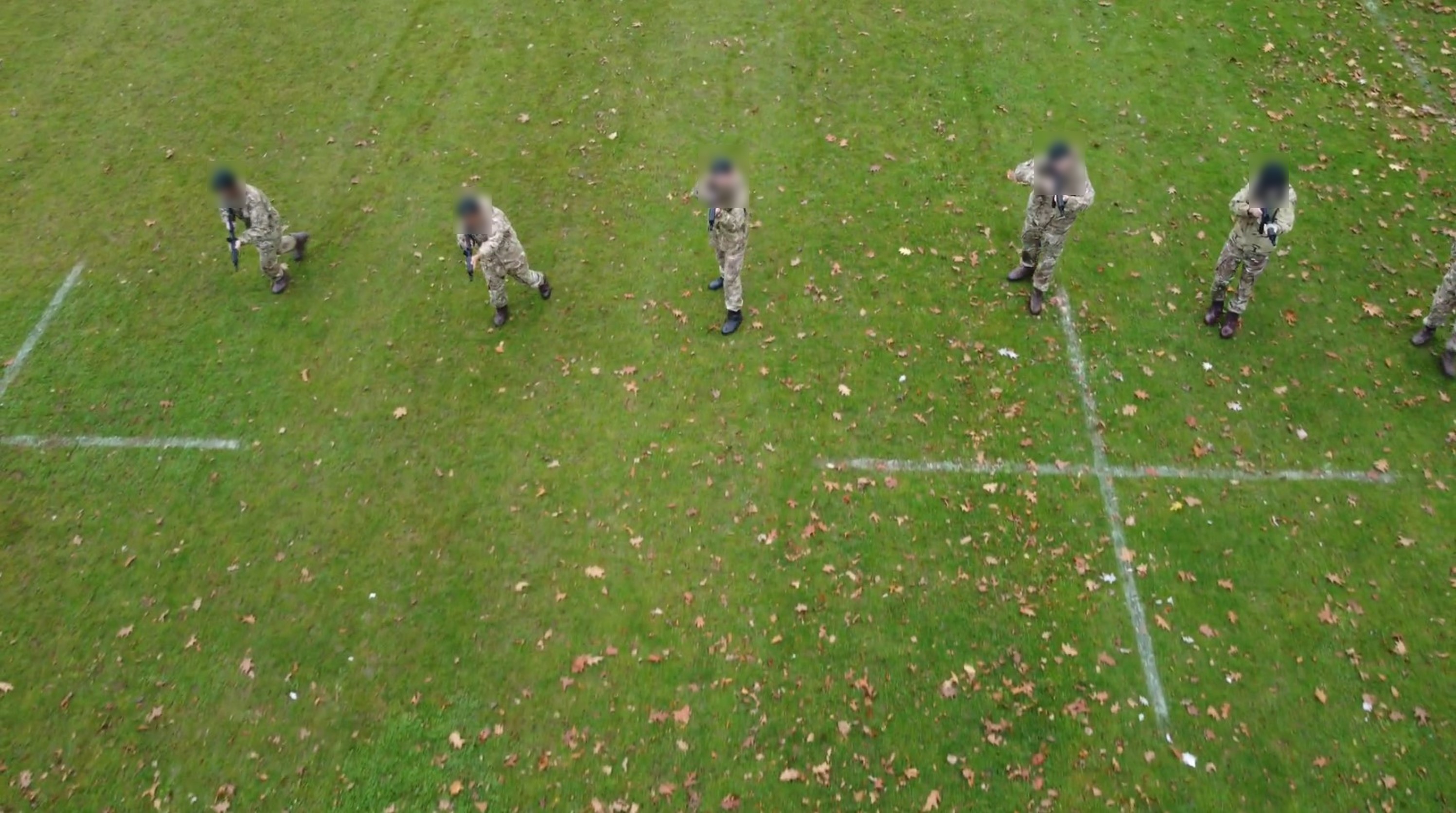}
    }\hfill
    \subfloat[Walk--Collapse\label{fig:scene_f}]{
        \includegraphics[width=0.152\textwidth]{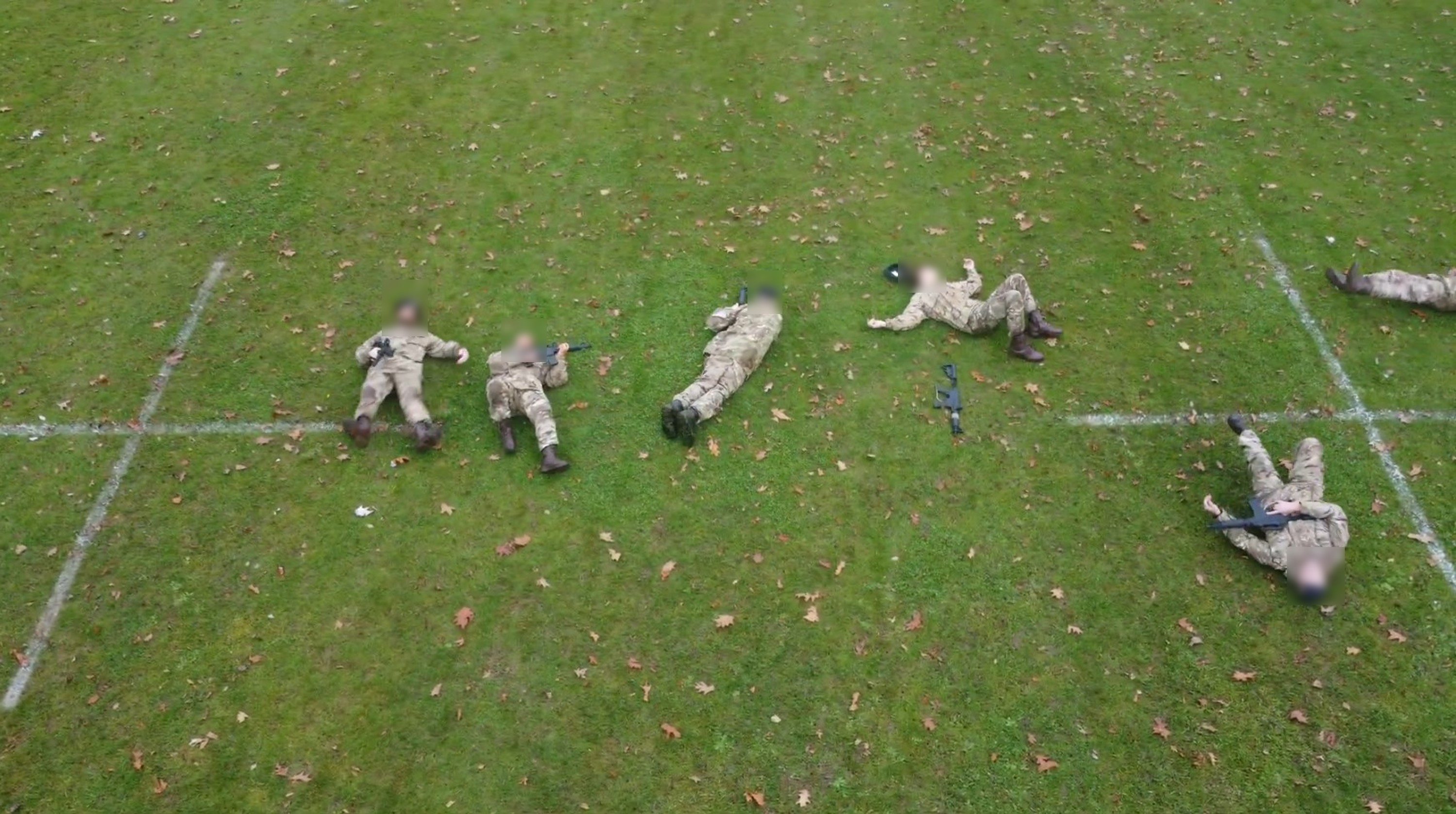}
    }\hfill
    \subfloat[Crawling\label{fig:scene_b}]{
        \includegraphics[width=0.152\textwidth]{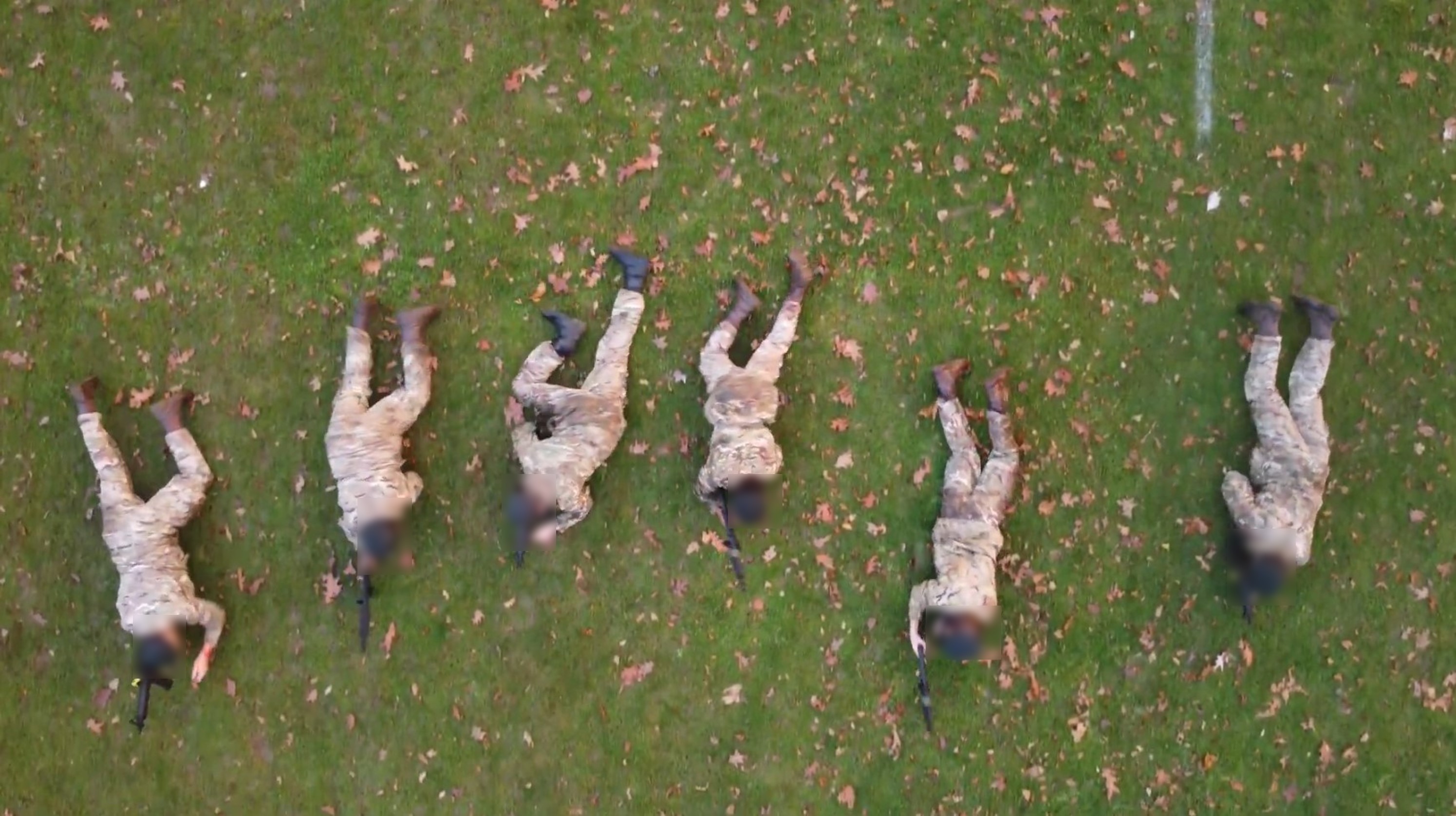}
    }\hfill
    \subfloat[Running\label{fig:scene_e}]{
        \includegraphics[width=0.152\textwidth]{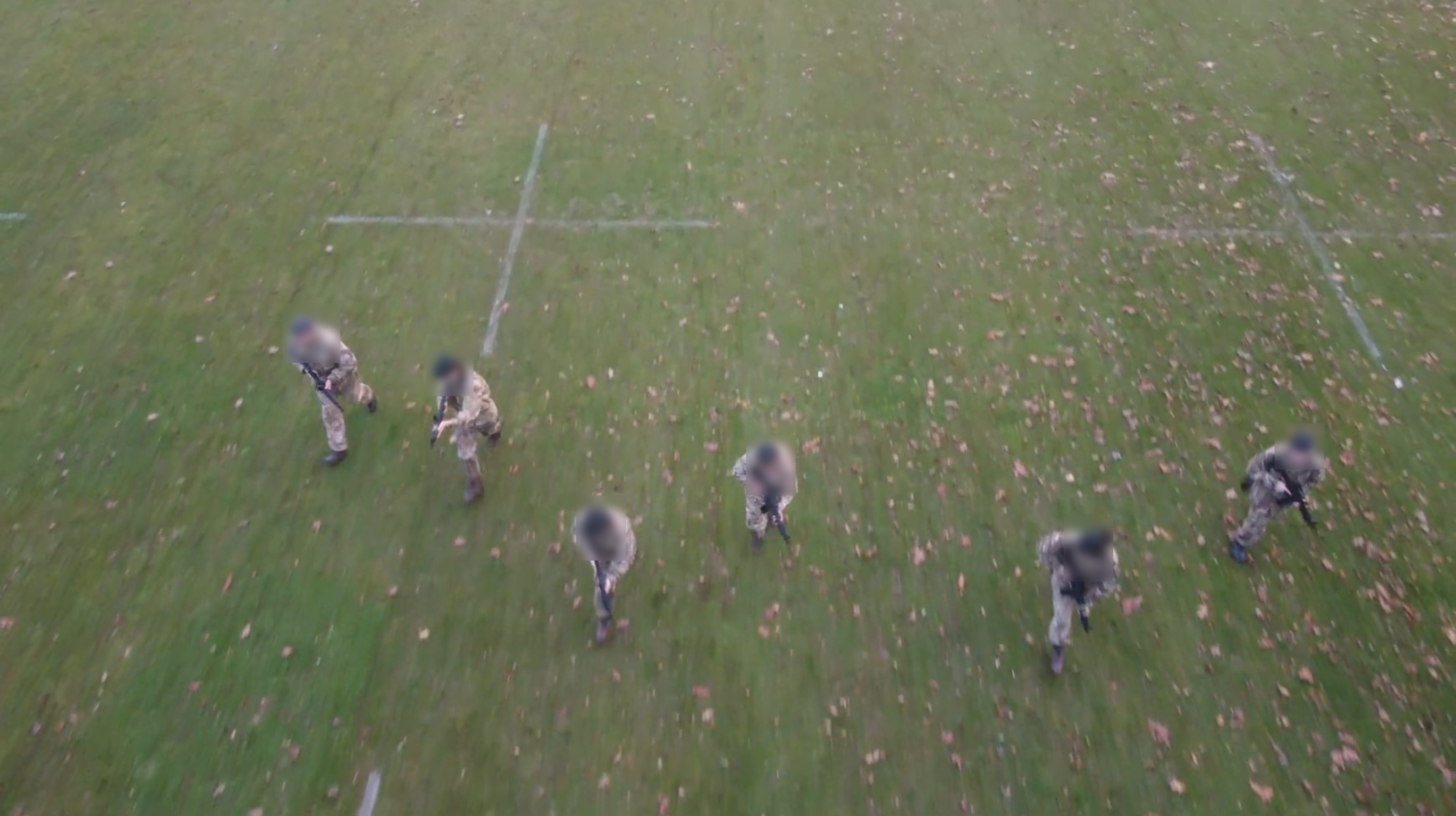}
    }
    \caption{Representative frames of six different choreographed actions, performed by cadets during data acquisition by drone.}
    \label{fig:drone_footage}
\end{figure*}

% \vspace{-0.5cm}
\subsection{Data Collection} \label{preliminaries}
% \subsection{Data Collection Setup}

We collected video and sensor data to support the development of ATRACT triage system. To this end, we arranged a field data capture activity with young healthy cadets. A high-level understanding of data capturing setup is shown in Fig.~\ref{fig:data_collection_and_sensors}. While performing activities, cadets were equipped with non-invasive Zephyr BioModule \cite{zephyrtechnology_2012_bioharness} and a GPS. BioModule is a physical-performance sensor for recording vital signals (heart rate, breathing rate, body posture, movement) and transmits signals through ECHO gateway Transceiver at 2.4 GHz (IEEE 802.15.4 connectivity) without signal attenuation within 300 yards. %The sensor data quality was ensured by monitoring through PSM (Physiological Status Monitoring) Training system.
With \emph{individual's consent}, participants were asked to perform a set of scripted activities, to imitate the behaviour of a wounded soldier with suffering from certain injuries (e.g., arm, head, or limping, etc.). Explicitly, individuals were asked to perform following actions in group, \textbf{injured:} (1) arm injury, (2) head injury, (3) limping, (4) walk and collapse, and \textbf{un-injured:} (5) crawling, and (6) running, shown in Fig. \ref{fig:drone_footage}. %While performing activities, cadets were equipped with non-invasive Zephyr BioModule \cite{zephyrtechnology_2012_bioharness} (Fig. \ref{fig:sensors_used}) and a GPS. BioModule is a physical-performance sensor attached to participant’s chest strap, to record vital signals (heart rate, breathing rate, acceleration, body posture) and transmit signals to the base station through ECHO gateway Transceiver at 2.4 GHz (IEEE 802.15.4 connectivity) without signal attenuation within 300-yards. The sensor data quality was ensured by monitoring through PSM (Physiological Status Monitoring) Training system. %The whole duration of video footage captured is about an hour, with participants performing different actions in different rounds.

\noindent
\textbf{MIMIC Database for data-realism gap:} We trained our generative model on MIMIC-I dataset~\cite{moody_data_1996} to augment the vital signs of healthy participants during data collection. MIMIC dataset contains records from over 90 ICU patients with periodic readings of their vital signs (heart and breathing rates, ECG signals, etc.). %, showing metrics such as heart and breathing rates, ECG signals, blood pressure, etc. %The clinical conditions available in this dataset are: Angina, Bleed, Brain Injury, Congestive Pulmonary, Failure/Pulmonary Edema, Cardiac Arrest, Cardiogenic Shock, Post-OP Coronary Artery Bypass Graft (Post-OP CABG), Post-OP Valve, Renal Failure, and Respiratory Failure.
Due to data realism gap and the ethical difficulty of capturing real injury vital signs, MIMIC dataset is the closest option that allows to demonstrate the adjustment of vital signs for intended clinical conditions. MIMIC dataset may also have additional measurements (e.g., systolic/diastolic/mean arterial blood pressure (ABP), Pulse etc.) for a subject; but, it was ensured that heart and breathing rate must be in the records for data augmentation.

\vspace{-0.11cm}
\subsection{Sensor Data Processing}
After data collection, it was noticed that deteriorating vital signs of actions (collapse, arm, head injury) can benefit from data augmentation, as the real-injury data acquisition for these actions is hard. %However, other actions (limping, crawling, running) are less reflective to injured vital signs.

% After the sensor data collection, it is noticeable that the vital signs of some of the actions (arm injury, head injury, collapse) can benefit from being adjusted, as the real-injury data acquisition for these three actions is hard. 
%After the sensor data collection for multi-modal triage, it is noticeable that deteriorating vital signs of actions (arm injury, head injury, collapse) can benefit from \emph{data augmentation}, as the real-injury data acquisition for these three actions is hard. However, other actions (limping, crawling, running) are less reflective to injured vital signs, and there is no need for data augmentation.

% For sensor data stream, a key challenges lies the sensor data acquisition for injured soldiers pertaining to the battlefield situation, therefore, we attempt to generate sensor data for injured soldiers by adjusting the vital signs of healthy participants in our data collection. To build an accurate triage model, it is necessary to construct a dataset capable of training, and one of the main limitations lies is the lack of relevant data due to the specific nature of the problem being studied in this paper. Therefore, it is crucial to build a generative model that can mimic the participants’ injury-related vital signs of distress. We devise this approach to augment the sensor data with video input for training the multi-modal pipeline.

\subsubsection{Sensor Data Augmentation}
\begin{figure*}[t]
    \centering

    % Left: architecture
    \begin{minipage}[t]{0.37\textwidth}
        \centering
        \vspace{0pt}
        \includegraphics[width=\linewidth]{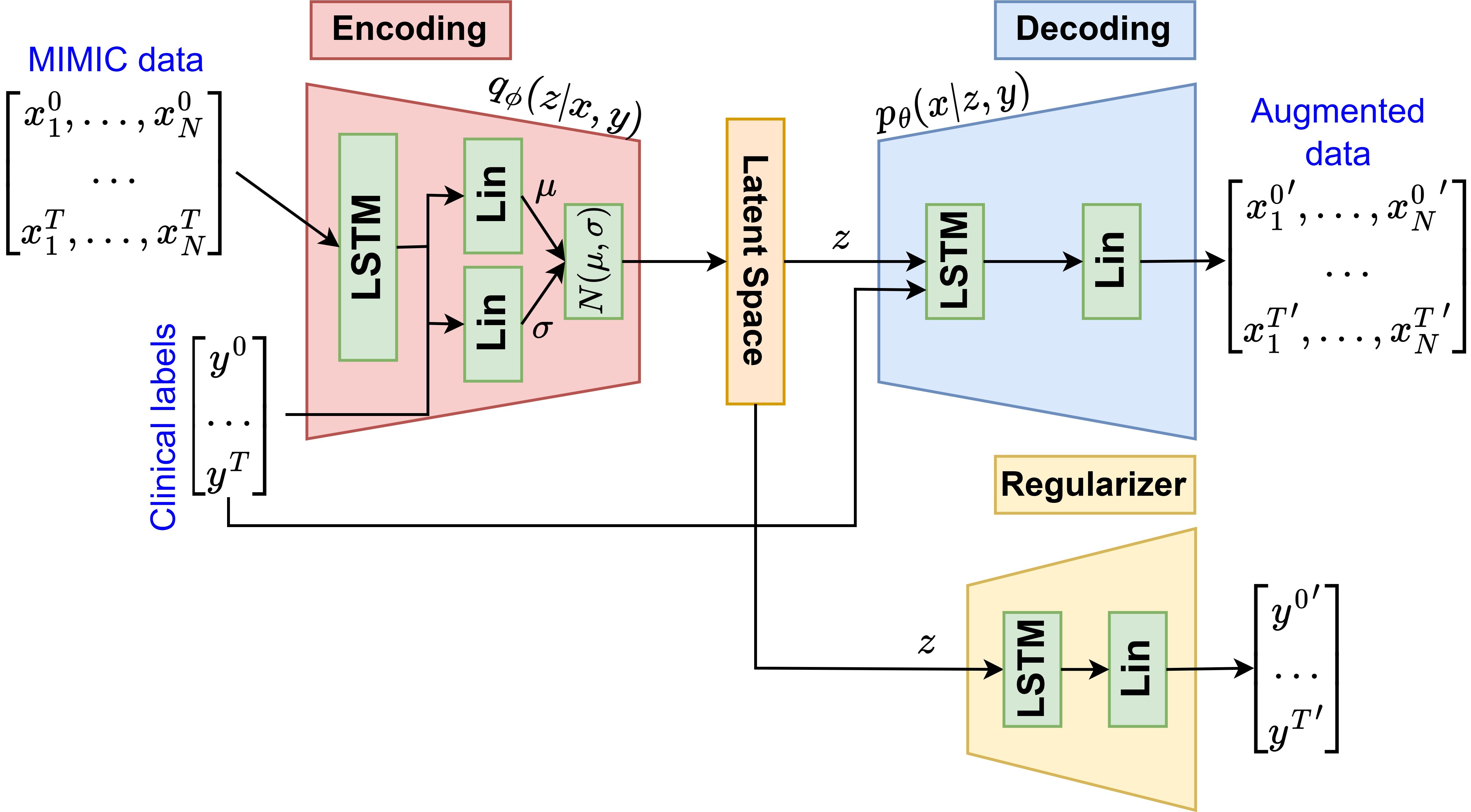}

        % \vspace{2pt}
        \small (a) CVVitAE architecture.
    \end{minipage}
    \hfill
    % Right: heatmaps
    \begin{minipage}[t]{0.62\textwidth}
        \centering
        \vspace{0pt}
        \includegraphics[width=0.35\linewidth]{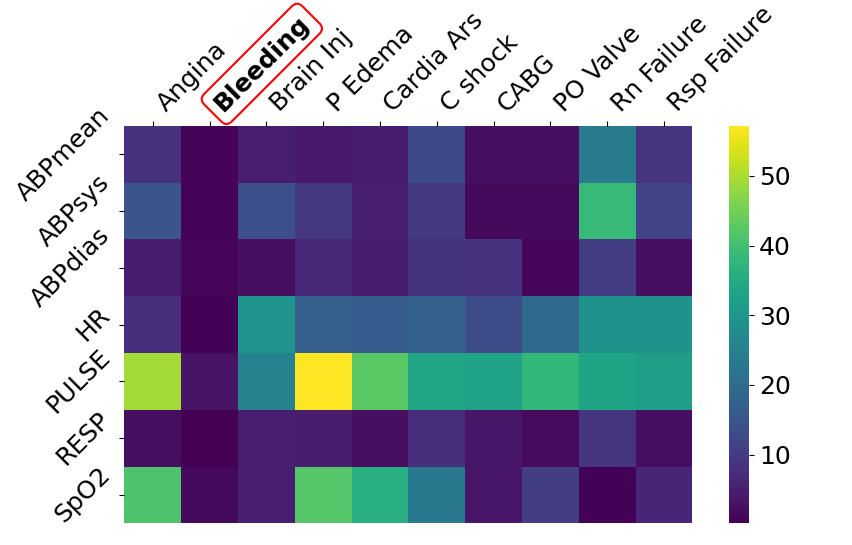}%\hfill
        \includegraphics[width=0.35\linewidth]{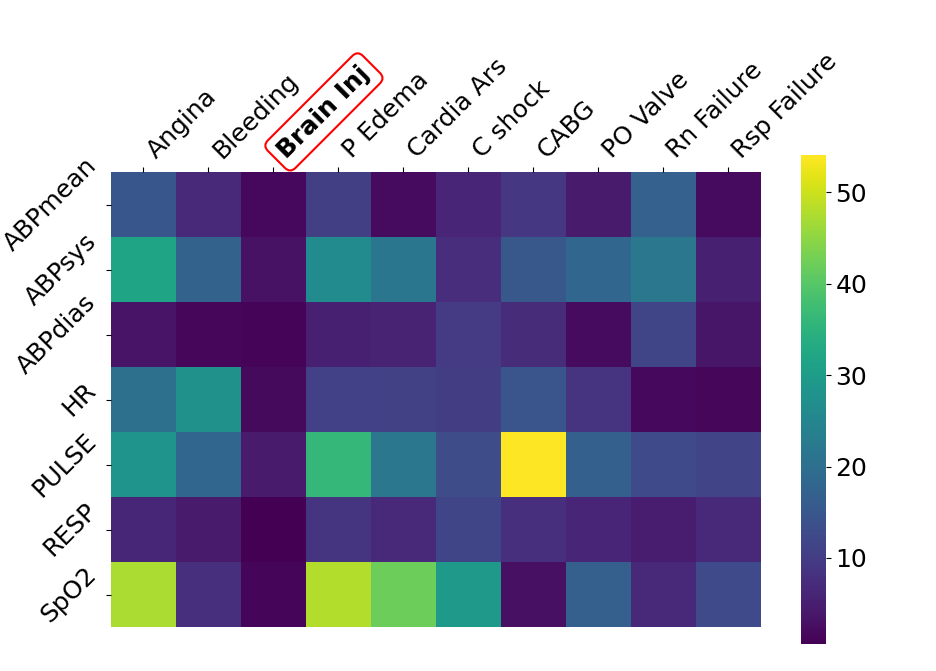}%\hfill
        \includegraphics[width=0.35\linewidth]{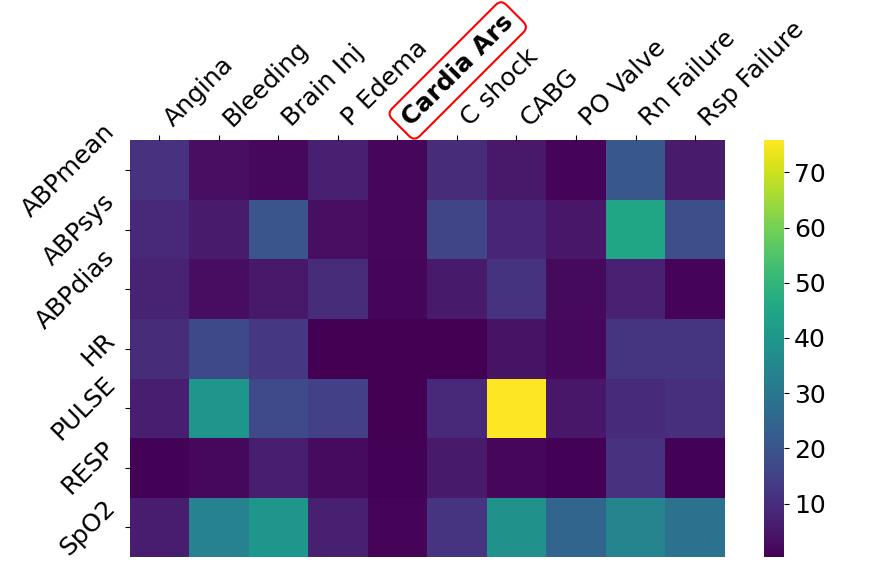}

        \vspace{3pt}
        \small (b) Proximity mapping between augmented data labels and MIMIC clinical labels.
    \end{minipage}

    \caption{CVVitAE architecture for vital-sign augmentation (arm injury, head injury, collapse), and validation through proximity mapping.}
    \label{fig:cvvitae_mapping_combined}
\end{figure*}

We employ a generative model to learn the hidden dynamics of MIMIC dataset and adjust the vital signs of our participants to better align with the conditions, being portraying in drone-captured videos. Inspired by~\cite{Sohn2015}, we devise \emph{CVVitAE} (Conditional Variational Vital Auto Encoder) for reshaping the data of healthy participants so that we can simulate their vital signs to appear if were affected by a clinical condition. % \cite{cvvitae_2025}.
The CVVitAE architecture (Fig.~\ref{fig:cvvitae_mapping_combined}a) condition the MIMIC vital signs to their clinical labels while passing through encoder $q_{\phi}(z|x,y)$ to form a latent space $z$. Then samples drawn from $z$ are passed to decoder $p_\theta(x|z,y)$ with their labels. The model is trained to minimise loss:

%In augmenting data, we employ a generative model to learn the hidden dynamics of MIMIC training dataset for adjusting the vital signs of our participants to better align with the conditions they are portraying in the drone-captured videos. Inspired by~\cite{kingma2013auto}, we devise \emph{CVVitAE} (Conditional Variational Vital Auto Encoder) for reshaping and augmenting the data of healthy participants so that we can simulate their vital signs to appear if were affected by a clinical condition \cite{cvvitae_2025}. The CVVitAE architecture (Fig.~\ref{fig:cvvitae_mapping_combined}a) includes \emph{LSTM layers} in encoder and decoder, followed by linear layers to learn the latent space distribution using mean ($\mu$) and standard deviation ($\sigma$). The model samples from the latent space to feed a decoder to map the outputs back to their original shape. Additionally, a \emph{regularizer} aligns the latent space to the intended data labels.

% After pre-processing (Section~\ref{sec:mimi_preparation}), we feed MIMIC-I data to the CVVitAE model to learn the relationships between vital signs and their corresponding clinical labels. 
%Following~\cite{kingma2013auto}, we condition the MIMIC vital signs to their clinical labels while passing through encoder $q_{\phi}(z|x,y)$ to form a latent space $z$. Then samples drawn from $z$ are passed to decoder $p_\theta(x|z,y)$ with their labels. The model is trained for the condition of minimising ELBO loss:
\vspace{-0.2cm}

\begin{equation}\label{eq:elbo_loss}
    \mathcal{L_\text{A}} = \mathbb{E}_{q_{\phi}(z|x,y)}[\log p_\theta(x|z,y)] - D_\text{KL}[q_{\phi}(z|x,y) \,||\, p_{\theta}(z|x)].
\end{equation}
We approximate the reconstruction term for minimizing the error:
$ \frac{1}{T}\sum_{t=1}^T (y'_t - y_t)^2$,
and use KL-divergence to normalize the distribution $\mathcal{N}(0, I)$ as regularisation.

The learning capacity of CVVitAE is strengthened by introducing a latent-space regulariser, which also uses an LSTM layer that receives the samples from latent space and attempts to map their respective clinical labels, using an objective function: $\mathcal{L_\text{B}} = -\sum_{i = 1}^C y_i \log\, p(y_i)$, where $C$ is the clinical labels and $p$ is the predicted labels distribution. Combining with loss in Eq.~\eqref{eq:elbo_loss}, the overall loss function becomes:
$\label{eq:general_loss}
\mathcal{L} = \alpha \times \mathcal{L_\text{A}} + \mathcal{L_\text{B}}$,
where $\alpha$ controls their contribution.

\subsubsection{Adjusting Vital Signs}\label{sec:setup_cvvitae}
We input our trained CVVitAE model with the samples of healthy participants to augment the vital signs data of required actions (collapse, arm, head injury). % , in our collected data the participants perform a set of activities that represent certain behaviours.
For these actions, there is no exact label class in MIMIC on which our CVVitAE model is trained, however, such injury-related actions have inherent mapping with corresponding class categories in MIMIC dataset; e.g., ``arm injury'' maps to ``Bleeding''; ``walk-collapse'' maps to ``Cardiac Arrest''; ``head injury'' maps to ``Brain Injury''.
%Since these acted conditions are not directly present in the dataset that we used to train our CVVitAE, we map these to the labels in the MIMIC dataset in the following manner: "arm injury" is mimicked using "Bleed"; "walk - collapse" is mimicked using "MI Arrest" (Heart Attack); "Head Injury" is mimicked using "Brain Injury". 
Fig. \ref{fig:cvvitae_mapping_combined}b highlights \emph{distance mapping} between augmented labels and the MIMIC labels. 
%the accuracy of the proposed CVVitAE in adjusting the vital signs of the healthy individuals according to the intended clinical labels. 
%the plots show a \emph{proximity metric} that compares the generated samples to the real values of the input features for each clinical class. In other words, proximity metric calculates the average value of each feature in the original dataset and then computes a map of distances between this average and the average of the corresponding features in the augmented samples. This procedure is performed for each clinical class being augmented, resulting in $m_{j,c}$ that quantifies the proximity between generated and real MIMIC-I samples. Formally, this can be written as the plots show a \emph{proximity metric} comparing generated samples to real input features for each clinical class. 
It computes average of each feature in the original data and measures the distance to corresponding averages in augmented samples, producing $m_{j,c}$ a similarity measure between augmented and real MIMIC samples. Mathematically, 
\begin{equation}\label{eq:metric}
    m_{j,c} = \frac{\sum_{t=1}^T \left(x'_j(t) - \overline{x}_j \right)}{T},
\end{equation}
where $j$ is the index of each feature at timestep $T$, and $c$ clinical label, $c \in \{1, \dots, C\}$.

Using Eq.~\ref{eq:metric}, we observe in Fig.~\ref{fig:cvvitae_mapping_combined}(b) for each augmented class (e.g., arm injury) features are always closer (distance minimization) to the average values of features in the intended class of MIMIC dataset (i.e. bleeding). This observation is consistent across all three classes (collapse, arm injury, head injury), endorsing that our model is capable of augmenting samples of healthy individuals that are aligned with the intended clinical conditions. The remaining actions, (crawling, running, limping) do not represent evident signs of injuries and their raw recordings generate better results without augmentation.  

\vspace{-0.2cm}

\subsection{Video Data Processing}
In ATRACT system (Fig.~\ref{fig:final flow}), visual stream captures the motion patterns (e.g., posture and body pose spatio-temporal movement) that vividly emerge when participants performed some actions. After video acquisition, we focus on transforming raw drone-captured footage into meaningful discriminative visual embeddings for multi-modal fusion. %Compared to sensor input, video features provides richer contextual and behavioural information for all actions.

\vspace{0.05cm}
\subsubsection{Bounding Box Extraction} 
Since multiple participants are present in each video frame, person-level localisation is a critical pre-processing step before any spatio-temporal modeling is performed. Therefore, we employ YOLO-v12~\cite{tian2025yolov12} detector to track each individual and extract their bounding boxes over the time. Additionally, drone footage is predominantly captured from a top- or oblique view, substantially differing from the terrestrial view common in standard object-detection datasets. As a result, a vanilla YOLO-v12 pre-trained on generic datasets exhibits degraded performance, particularly in cases of partial-self occlusion, overlapping participants, and scale variation. To address, we fine-tuned YOLO-v12 on our soldier-dataset that explicitly reflects the aerial view of subject, group formations, and  occlusion. This task-specific adaptation leads to more reliable person detection, temporally consistent bounding boxes, and robust tracking of individuals across the consecutive frames. Each detected bounding box corresponds to a single participant, and the resulting sequence of bounding boxes over time forms a person-centric video clip, which is fed to encoder for spatio-temporal feature extraction.

% \begin{figure}[t]
%     \centering
%     \subfloat[Raw footage\label{fig:scene_a}]{
%         \includegraphics[width=0.47\columnwidth]{resources/Picture1.png}
%     }\hfill
%     \subfloat[Masked frames\label{fig:scene_b}]{
%         \includegraphics[width=0.47\columnwidth]{resources/Picture2.png}
%     }\hfill
%     \caption{Person detection in raw footage using fine-tuned YOLOv12 and mask extraction.}
%     \label{fig:drone_footage2}
% \end{figure}

% Since there are more than one subjects in each frame, first of all movement of each person is tracked and Bounding Box (BBox) is extracted for each subject using finetuned YOLO-v12 model. Vanilla YOLO-v12 did not perform well due to the top-view of the subjects, therefore, finetuning on our soldier dataset improved the detectors performance and led to better BBox extraction for each subject in the frame. 

\vspace{0.05cm}
\noindent
\subsubsection{Data Synchronization}  
Another major challenge emerged during multi-person tracking, when a soldier exits and re-enters the field-of-view, or when participants overlap/swap positions and their bounding box IDs changed. This led to an \emph{unsynchronised bounding box sequences}, breaking the temporal continuity required for associating each individual with their corresponding temporal sensor reading.

To resolve this, we synchronised the frames by re-ordering and matching bounding boxes using spatial proximity, trajectory continuity, and temporal smoothness. This ensured that each subject maintained a unique identity throughout the sequence, enabling reliable alignment between video crops and sensor data. Without correction, sensor input would be incorrectly paired with visual trajectories resulting in poor performance. Following detection and synchronization, person-centric crops were extracted as:

\begin{equation}
\texttt{Video}_i =
\left\{
\begin{aligned}
&\texttt{person}_1 : [f_1, \dots, f_T], \\
&\hdots \\
&\texttt{person}_N : [f_1, \dots, f_T]
\end{aligned}
\right.
\label{eq:vide_eq}
\end{equation}

\noindent
Above eq.~\ref{eq:vide_eq} guarantees one-to-one mapping between visual and sensor input, offering a practical solution and important perspective for clinically meaningful triage support.

\subsubsection{Video Encoder}
After extracting the trajectories of each individual using bounding box tracking, visual stream focuses on learning discriminative spatio-temporal motion representations that characterise the movement over time for each participant. For this purpose, cropped frames of each individual are stacked to form a three-dimensional spatio-temporal cube, where the spatial dimensions encode body posture and orientations, while temporal dimension captures the evolution of motion dynamics across the frames. This formulation allows the model to emphasise motion patterns rather than static appearance, which is particularly important for distinguishing injury, (limping, crawling, collapsing) that may share similar visual appearance in isolated frames but prominently differ in their temporal progression. The resulting spatio-temporal cubes are processed by a lightweight two convolutional layered video encoder, each followed by a max-pooling operation. The convolutional layers progressively extract salient spatial features, while max-pooling reduces spatial resolution, improving robustness to subtle viewpoint variations introduced by drone movement, and also limits the computational complexity. This design captures both fine-grained motion cues and global behavioural trends over time, ultimately transforming the input cube into a fixed-length visual feature vector. This compact representation serves as visual embedding used for multi-modal learning; meanwhile, input-output feature dimension at each layer is detailed in the supplementary section. %Table~\ref{tab:feature_dimensions_detailed}.

\vspace{-0.2cm}

\subsection{Multi-modal Pipeline}
A multi-modal pipeline to jointly exploit the complementary strengths of visual and sensor inputs is designed. While visual stream captures motion cues from drone videos, the complementary sensor stream reflects the participant’s physical condition, operating two modalities independently for feature encoding and are subsequently fused for unified representation.

\begin{figure}[!t]
    \centering
    \includegraphics[width=0.95\columnwidth]{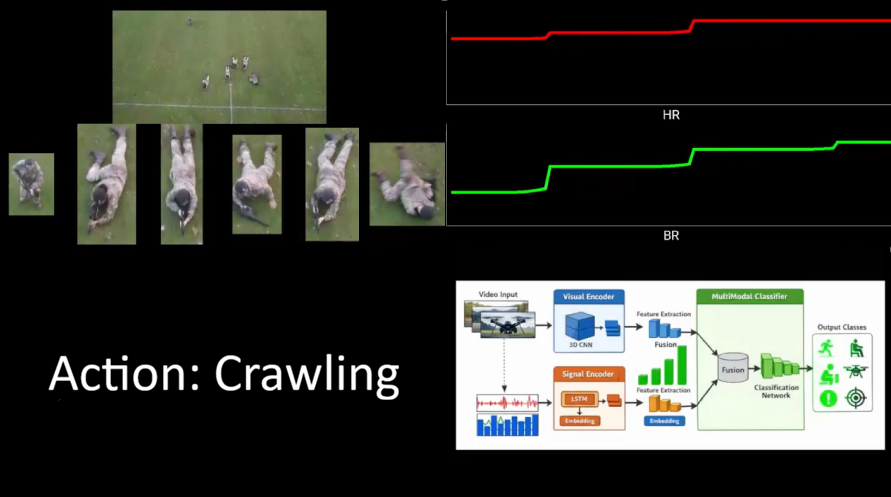}
    \caption{Graphical Interface of ATRACT system, which integrates video input with sensor modality (HR, BR, etc.) for multi-modal learning and decision support of casualty triage.}
    \label{fig:gui}
\end{figure}

Specifically, visual encoder generates a $512$-d fixed-length embedding that summarises the spatio-temporal evolution of each participant’s movement by aggregating frame-level motion over time. In parallel, sensor stream is processed using LSTM encoder that models temporal dependencies in physiological signals and outputs a compact $128$-d feature embedding corresponding to participant’s health state. Video and sensor modalities are asynchronous and operate at different temporal rates. Therefore, data synchronisation is required for an effective alignment between visual and sensor features for fusion and multi-modal learning. This concatenated feature vector is then projected onto a fully connected layer for classification. A non-linear activation and dropout layer for regularisation are applied to improve the generalisation and reduce overfitting. Finally, a classification layer maps features to class logits of the target class. A simple graphical interface of ATRACT system is shown in Fig.~\ref{fig:gui}, illustrating how drone video, sensor input, and multi-modal fusion systematically work together for casualty triage.

% \textbf{Implementation details}
% We trained all models on a single NVIDIA A40 GPU (48\,GB) using the Adam optimiser with a learning rate of $1\times10^{-4}$. During training, standard augmentation included horizontal flipping and normalisation, with input resizing set to $128\times64$ for our CNN, $224\times224$ for S3D, and $112\times112$ for the other video backbones. For testing, only resizing and normalisation were applied using the same backbone-specific input resolutions.

\subsection{Communication and Active Sensing module}
Another key component is the communication and active sensing module, which enables continuous sensor data acquisition and reliable transmission from the field to base station. To this end, a dedicated \textit{Multi-modal Sensor} (MuS) is developed to integrate sensing, power supply, and wireless transmission within a compact portable unit. MuS is designed to continuously capture vital signs from body-worn sensors; thereby providing a continuous stream of physiological (signals) and behavioural (video) evidence during the battlefield triage. In parallel, drone serves as an active sensing platform, collecting complementary RGB video to observe external motion patterns. Two sensing modalities collectively support our core concept that casualty assessment should benefit from the joint contextualisation of external visual evidence and internal physiological state.

\begin{figure}[t]
    \centering
    \subfloat[\label{fig:scene_a}]{
        \includegraphics[width=0.3\columnwidth]{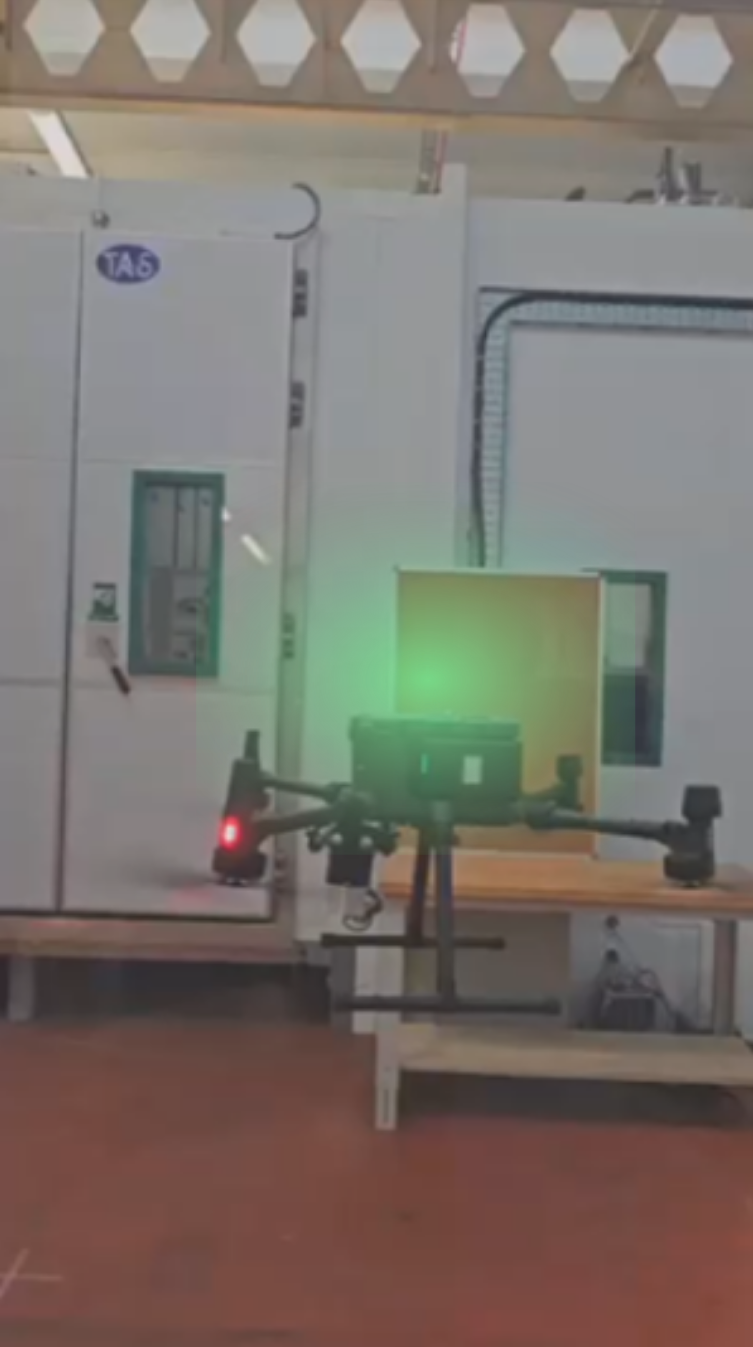}
    }\hfill
    \subfloat[\label{fig:scene_b}]{
        \includegraphics[width=0.3\columnwidth]{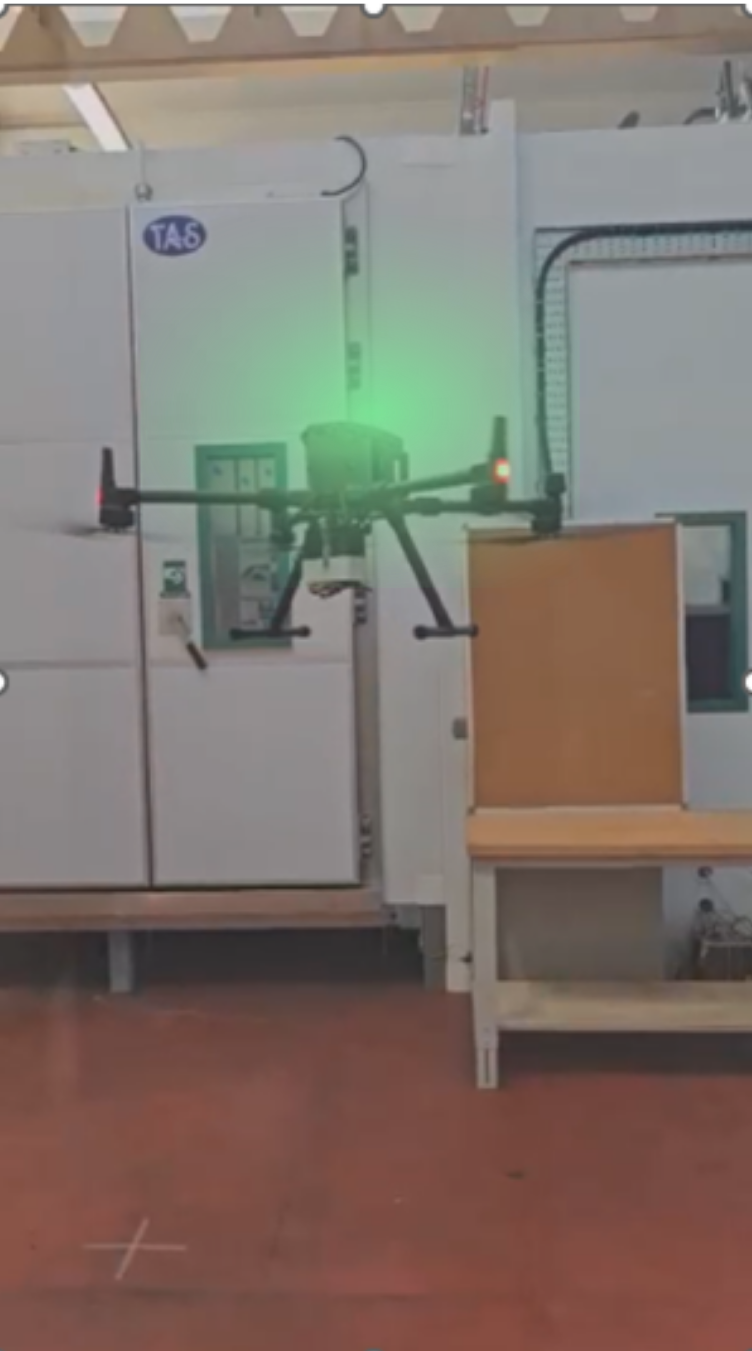}
    }\hfill
    \subfloat[\label{fig:scene_b}]{
        \includegraphics[width=0.3\columnwidth]{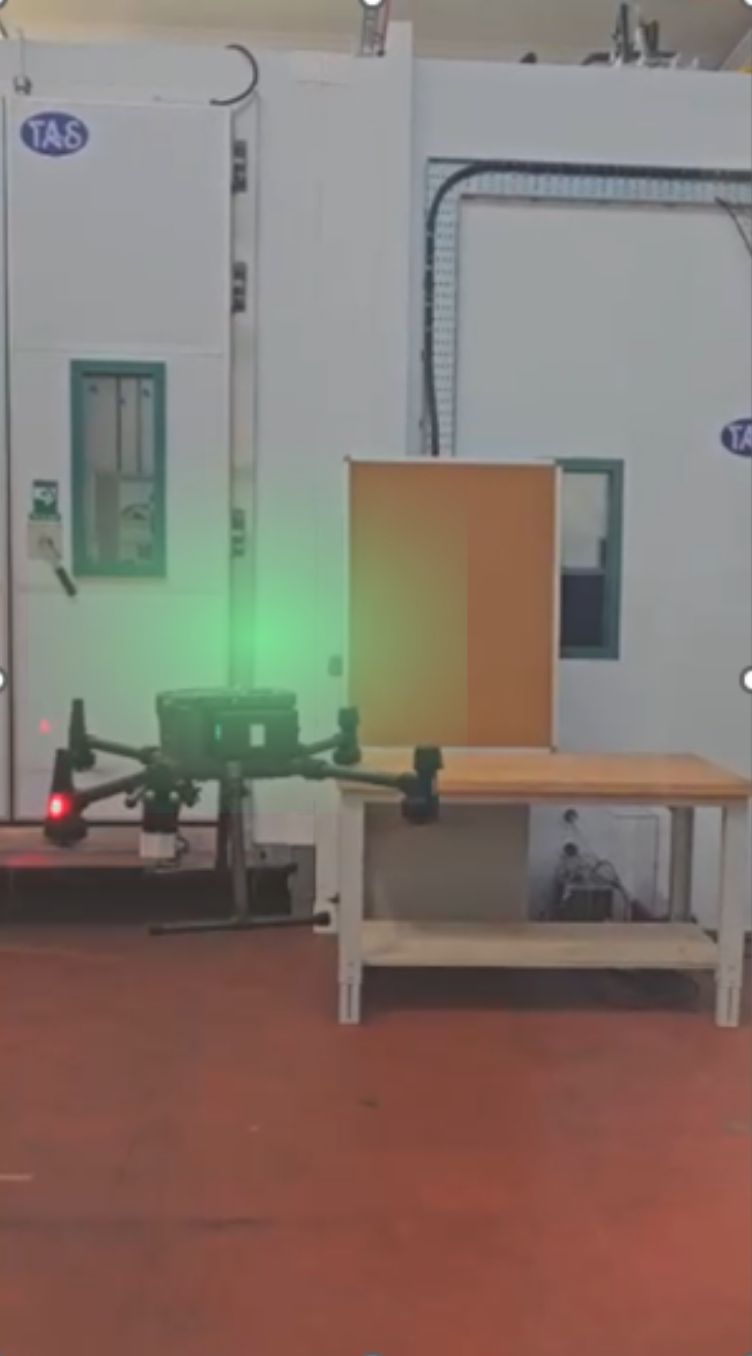}
    }\hfill
    \caption{Stable Drone flight while carrying the communication module.}
    \label{fig:khizer}
\end{figure}

To support this integration, an onboard \textit{data collection and transmission} (DaCT) system is required to synchronise data packets and transmit MuS readings together with drone-captured video to the base station for downstream multi-modal fusion and triage inference. This communication channel is designed with practical constraints in mind and integrated with drone system (Fig.~\ref{fig:khizer}) to ensure stable flight operation and reliable sensing. Such active sensing communication protocol is particularly important in the contested environments, where reliable and fast communication link can substantially improve remote casualty localization, better judge health state, and channelize the readiness of medics for triage support.

\section{Results}\label{Results_Sec2}
This section presents empirical evaluation of ATRACT system. Then, we examine class-wise accuracy for individual and different combinations of sensor modalities to figure out the most informative cues for reliable triage support.

% \vspace{-0.25cm}

\subsection{Multi-modal Data Fusion}
% We compare the effect of late fusion for different video backbones in Table~\ref{tab:sota_comparision}.

Table~\ref{tab:sota_comparision} shows that late fusion benefits differently across video backbones. Overall, R(2+1)D achieves the strongest performance, while R3D and our lightweight CNN benefit from sensor data augmentation; in contrast, S3D and MC3 remain unchanged.
Without sensor data augmentation, the model achieves 71.4\% for S3D, MC3, and our CNN; while drops to 57.1\% for R3D, and reaches \textbf{85.7}\% for R(2+1)D. After augmentation, performance remains unchanged for S3D and MC3, improves from 57.1\% to 71.4\% for R3D, and increases from 71.4\% to \textbf{85.7}\% for our CNN, while R(2+1)D remains at \textbf{85.7}\%. Since augmentation affects only the sensor stream, these changes indicate how effectively each video encoder exploit more realistic and class-informative physiological cues during fusion.

\begin{table}[!htbp]
\caption{Performance of late fusion for video and wearable sensor inputs (HR + BR + BP + Move.) with (\textbf{w}) and without (\textbf{w/o}) data augmentation for recognizing actions. Results are reported as classification accuracy (\%). Best in \textbf{Bold}.}
\centering
\resizebox{\columnwidth}{!}{
\setlength{\tabcolsep}{14pt}
\renewcommand{\arraystretch}{1.35}
\begin{tabular}{lccccc}
\toprule
\textbf{Modalities} & \textbf{S3D} & \textbf{R3D} & \textbf{MC3} & \textbf{R(2+1)D} & \textbf{Ours}\\
\midrule
Video + Sensor (w/o) & 71.4 & 57.1 & 71.4 & \textbf{85.7} & 71.4 \\
Video + Sensor (w) & 71.4 & 71.4 & 71.4 & \textbf{85.7} & \textbf{85.7} \\
\bottomrule
\end{tabular}
\label{tab:sota_comparision}
}
\end{table}

It is noticed that sensor data augmentation is helpful when visual stream alone is either noisy or less informative for drone video processing. The gain for R3D and, more notably, our lightweight CNN suggests that remapping of raw vital signals to injury-related realistic sensor signals provides complementary evidence that the video branch does not fully capture on its own. This is plausible because some target actions are visually ambiguous in drone footage, while heart rate, breathing rate, posture, and movement provide additional information about the underlying physical state. In contrast, the unchanged result for R(2+1)D suggests that its factorised spatio-temporal design already extracts sufficiently discriminative motion cues, so sensor data realism does not further improve performance under current settings.

A practical finding is that our two-layer CNN matches the best accuracy of \textbf{85.7}\% once paired with augmented sensor data. This suggests that a lightweight visual model can still support effective triage assessment when complemented by realistic physiological signals, offering a practical alternative to relying solely on a heavier pre-trained video backbone. 
% This suggests that, for this task, effective multi-modal fusion for a light weight model can benefit from the data realism gap which will be an alternative to using a heavier pre-trained video backbone. 
Given that large video models are typically pre-trained on ground-captured action datasets (e.g., Kinetics), whereas ATRACT operates on aerial footage with occlusion, clutter, and viewpoint variation, a light visual encoder adapting more readily to the target domain can be a strong choice when supported with informative physiological sensor inputs. %This interpretation is further supported by the learning curves in Fig.~\ref{fig:Train-testing curves}, showing rapid fitting to the training data but less stable test performance, which reflects the difficulty of generalisation on a small and visually challenging dataset.

\vspace{0.07cm}

\noindent
\textbf{Implementation details}
We trained all models on a single NVIDIA A40 GPU (48\,GB) using Adam optimiser with learning rate $1\times10^{-4}$. Sensor data is represented as a time-series matrix of shape $[T, D]$, with $T$ timesteps for $D$ modalities (e.g., heart, breathing rate, body posture, and movement). During training, standard augmentation included horizontal flipping and normalisation, with input resize to $128\times64$ for our CNN, $224\times224$ for S3D, and $112\times112$ for other video backbones. For testing, only resizing and normalisation were applied for the same backbone-specific input resolutions.

\vspace{-0.3cm}

\subsection{Ablation Study}
% \vspace{-0.075cm}

\begin{figure}[!htbp]
    \centering
    \subfloat[CNN backbone\label{fig:scene_a}]{
        \includegraphics[width=0.48\columnwidth]{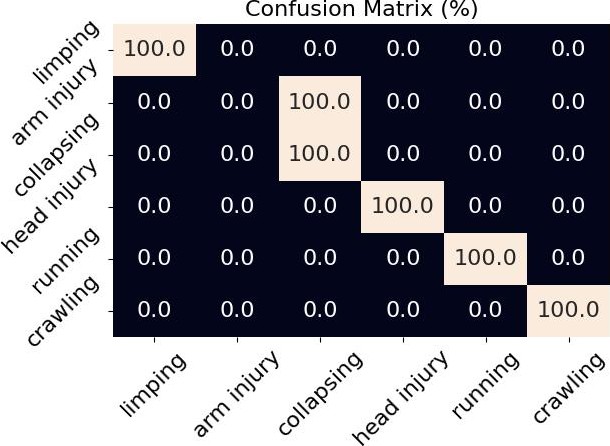}
    }
    \subfloat[R(2+1)D backbone\label{fig:scene_b}]{
        \includegraphics[width=0.48\columnwidth]{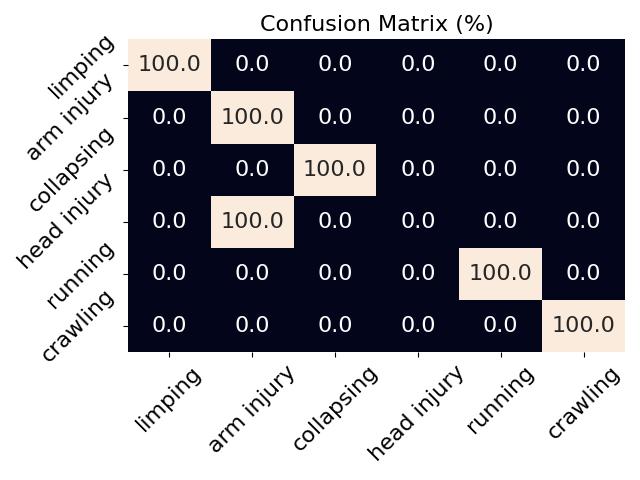}
    }
    \caption{Confusion matrix (\%) for our CNN and R(2+1)D backbone.}
    \label{fig:confusion_matrix}
\end{figure}

\paragraph{Class-wise Accuracy}
In Fig.~\ref{fig:confusion_matrix}, class-wise accuracy in confusion matrices shows that both backbones recognise \textit{limping}, \textit{collapsing}, \textit{running}, and \textit{crawling} with perfect accuracy, indicating that these classes exhibit sufficiently distinctive spatio-temporal patterns for reliable discrimination in drone videos. However, both models struggle in handling the arm and head injury. Our 2-layer CNN correctly identifies \textit{head injury} but misclassifies \textit{arm injury} as \textit{walking/collapsing}, suggesting its lightweight representation is effective for capturing global motion irregularities but is less sensitive to upper-body injury cues when overall motion patterns resemble with collapse. In contrast, R(2+1)D correctly recognises \textit{arm injury} but confuses \textit{head injury} with \textit{arm injury}, which implies that its factorised spatial-temporal modeling is better at preserving subtle local motion cues, yet may over-associate visually similar injured postures. 
Overall, both backbones are strong on the remaining classes, while the confusion between semantically similar arm injury and head injury reinforces that video evidence alone can be ambiguous in challenging drone views and multi-modal fusion is important for more reliable triage support.

% Overall, both backbones are strong on other classes, where confusion lies in semantically close arm and head injuries, reinforcing our hypothesis that video evidence alone can remain ambiguous in challenging drone views and multi-modal input fusion is necessary to improve the robustness and injury-specific discrimination.

% \begin{table}[!htbp]
% \caption{Classification Accuracy (\%) for each action.}
% \centering
% \resizebox{\columnwidth}{!}{
% \setlength{\tabcolsep}{14pt}
% \renewcommand{\arraystretch}{1.35}
% \begin{tabular}{lccc|cc}
% \toprule
% \multicolumn{4}{c|}{\textbf{Injured $\sim$ 75\%}}  & \multicolumn{2}{c}{\textbf{Un-injured $\sim$ 100\%}}\\
% \midrule
% \textbf{Limping} & \textbf{Arm Injury} & \textbf{Walk-collapse} & \textbf{Head Injury} & \textbf{Running} & \textbf{Crawling}\\
% \midrule
% 100 & 100 & 100 & 0 & 100 & 100 \\
% \bottomrule
% \end{tabular}
% \label{tab:class-wise accuracy}
% }
% \end{table}
%%%%%%%%%%%%%%%%%%%%%%%%%%%%%%%%%%%%%%%%%%%%%
% \begin{figure}[!htbp]
%     \centering
%     \subfloat[CNN backbone \label{fig:scene_a}]{
%         \includegraphics[width=0.52\columnwidth]{resources/train_test/cm_cnn_original_split4.jpg}
%     }
%     \subfloat[R(2+1)D backbone\label{fig:scene_b}]{
%         \includegraphics[width=0.45\columnwidth]{resources/train_test/cm_r21d_original_split4.jpg}
%     }
%     \caption{Confusion matrix (\%) for our 2-layer CNN and R(2+1)D backbone.}
%     \label{fig:confusion_matrix}
% \end{figure}
%%%%%%%%%%%%%%%%%%%%%%%%%%%%%%%%%%%%%%%%%%%%%%%%%

\noindent
\paragraph{Performance using different modalities}
Table~\ref{tab:individual_signal_ablation} provides a useful uni-modal baseline for interpreting the fusion results, because it separates the information content of video input from that of different sensor signals. The \emph{strongest} standalone result is the video-only setting at \textbf{71.4}\%, which is consistent with the role of visual stream in ATRACT as the primary source of spatio-temporal behavioural evidence from the aerial footage. However, Table~\ref{tab:individual_signal_ablation} also shows that the body-worn sensor channels are not equally informative when used in isolation. Among the sensor signals, body posture is the \emph{strongest single cue} at 57.1\%, followed by breathing rate at 42.9\%, whereas heart rate and movement alone are much weaker at 25.6\% and 28.6\%, respectively. This pattern is scientifically plausible in the context of casualty-action recognition: several target classes in ATRACT, such as limping, collapse, and head or arm injury, are expressed more directly through body configuration and overall physical state than through a single physiological variable. Table~\ref{tab:individual_signal_ablation}, therefore, \emph{reinforces} the paper’s main theme that no single modality is sufficient on its own, while also showing that posture is the most informative sensor input for this action recognition.

A second important observation in~Table~\ref{tab:individual_signal_ablation} is that performance depends more on \emph{complementarity} than on simply adding more signals. Combinations that include posture, such as HR + Posture, BR + Posture, and HR + BR + Posture, all reach \textbf{71.4}\%, matching the video-only baseline, whereas combinations without posture are generally weaker: HR + BR and BR + Move.\ remain at 57.1\%, while HR + Move.\ and HR + BR + Move.\ drop to 42.9\%. Even the full four-signal combination (HR + BR + BP + Move.\ ) does not improve beyond 57.1\%, indicating that additional modalities can introduce redundancy or noise rather than useful discrimination under a small and difficult dataset. This justification is aligned with other ablation results, where video fusion with posture gives the best single-modality gain, and where multi-signal fusion is shown to depend on specific combinations rather than the number of inputs. Taken together, these findings support a \emph{key argument} of the paper: in challenging drone views, video evidence alone can remain ambiguous for clinically similar injury states, but carefully selected sensor cues, especially posture, and to a lesser extent breathing rate, provide the most meaningful complementary information for more robust and trustworthy remote triage. Further results on video input with different combinations of sensor inputs are provided in the supplementary section. 

\begin{table}[t]
\caption{Ablation study of video-only and signal-only, combinations between signal modalities for casualty-triage action recognition. BR (Breathing rate), HR (Heart rate), BP (Body posture) and Move. (Body Movement). Best in \textbf{Bold}.}
\label{tab:individual_signal_ablation}
\centering
\footnotesize
\setlength{\tabcolsep}{3pt}
\renewcommand{\arraystretch}{1.08}
\begin{tabularx}{\columnwidth}{>{\raggedright\arraybackslash}X
                                >{\centering\arraybackslash}p{0.075\columnwidth}
                                >{\centering\arraybackslash}p{0.075\columnwidth}
                                >{\centering\arraybackslash}p{0.075\columnwidth}
                                >{\centering\arraybackslash}p{0.10\columnwidth}
                                >{\centering\arraybackslash}p{0.10\columnwidth}
                                >{\centering\arraybackslash}p{0.14\columnwidth}}
\toprule
\textbf{Setting} & \textbf{Video} & \textbf{HR} & \textbf{BR} & \textbf{Move.} & \textbf{Posture} & \textbf{Acc. (\%)} \\
\midrule
Video                    & $\checkmark$ & --            & --            & --            & --            & \textbf{71.4} \\
\midrule
HR                       & --           & $\checkmark$  & --            & --            & --            & 25.6 \\
BR only                      & --           & --            & $\checkmark$  & --            & --            & 42.9 \\
Move.                 & --           & --            & --            & $\checkmark$  & --            & 28.6 \\
BP                  & --           & --            & --            & --            & $\checkmark$  & 57.1 \\
\midrule
HR + BR                      & --           & $\checkmark$  & $\checkmark$  & --            & --            & 57.1\\
HR + Move.                & --           & $\checkmark$  & --            & $\checkmark$  & --            & 42.9 \\
HR + BP                 & --           & $\checkmark$  & --            & --            & $\checkmark$  & \textbf{71.4} \\
BR + Move.                & --           & --            & $\checkmark$  & $\checkmark$  & --            & 57.1 \\
BR + BP                 & --           & --            & $\checkmark$  & --            & $\checkmark$  & \textbf{71.4} \\
\midrule

HR + BR + BP            & --           & $\checkmark$  & $\checkmark$  & --            & $\checkmark$  & \textbf{71.4} \\
HR + BR + Move.           & --           & $\checkmark$  & $\checkmark$  & $\checkmark$  & --            & 42.9 \\
\midrule
HR + BR + BP + Move. & --           & $\checkmark$  & $\checkmark$  & $\checkmark$  & $\checkmark$  & 57.1 \\
\bottomrule
\end{tabularx}
\end{table}

\noindent
\section{ATRACT's Explainability and Trustworthiness} \label{label:explainability}
\begin{figure}[t]
    \centering
    \setlength{\tabcolsep}{1pt}
    \renewcommand{\arraystretch}{1.0}
    \resizebox{\columnwidth}{!}{%
    \begin{tabular}{@{}cccccc@{}}
        % Column labels above Row 1
        \scriptsize Arm injury & \scriptsize Head injury & \scriptsize Limping & \scriptsize Collapsing & \scriptsize Crawling & \scriptsize Running \\[0.4mm]

        % Row 1
        \includegraphics[width=0.155\columnwidth]{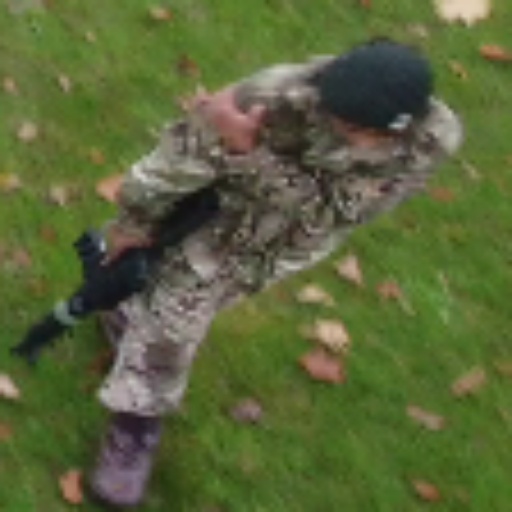} &
        \includegraphics[width=0.155\columnwidth]{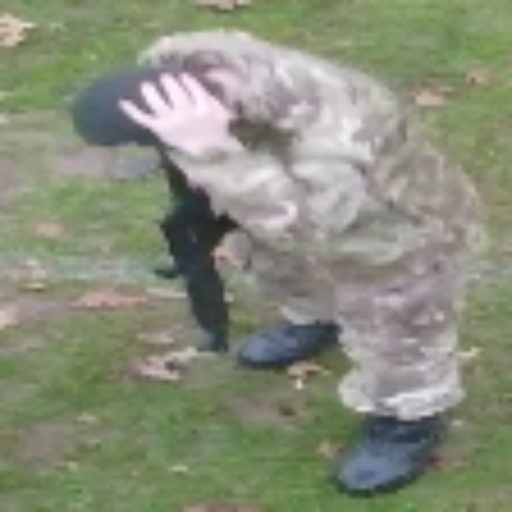} &
        \includegraphics[width=0.155\columnwidth]{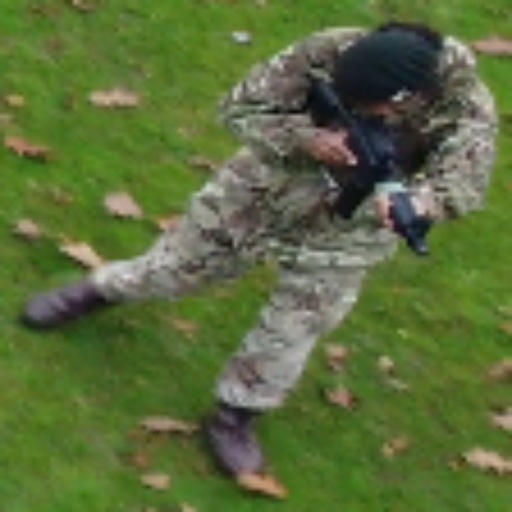} &
        \includegraphics[width=0.155\columnwidth]{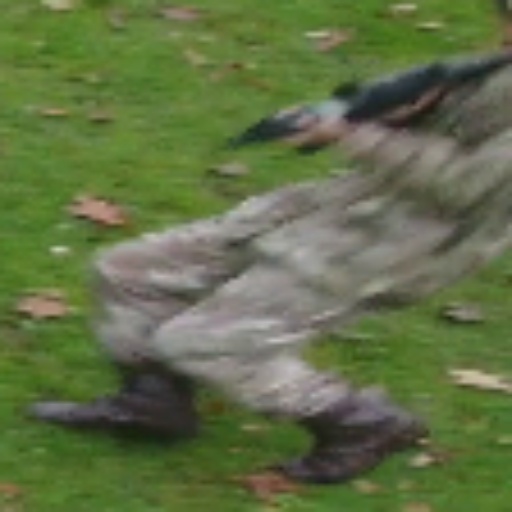} &
        \includegraphics[width=0.155\columnwidth]{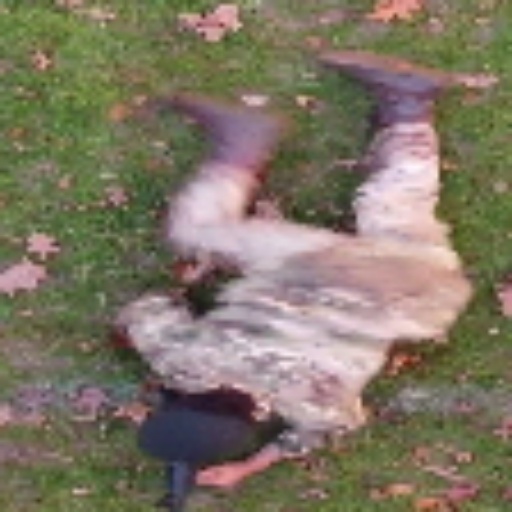} &
        \includegraphics[width=0.155\columnwidth]{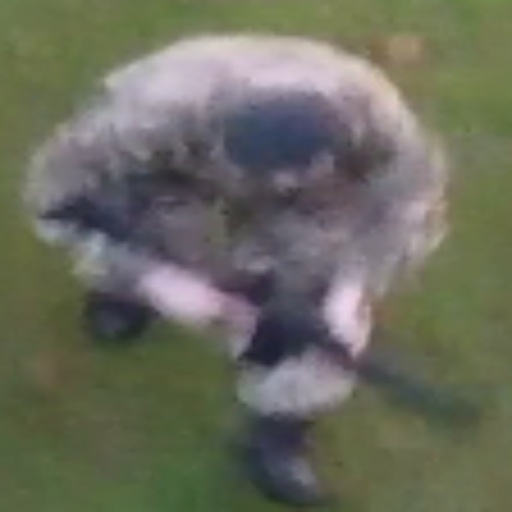} \\[-0.3mm]
        \multicolumn{6}{c}{\scriptsize (a) Image masks} \\[0.6mm]

        % Row 2
        \includegraphics[width=0.155\columnwidth]{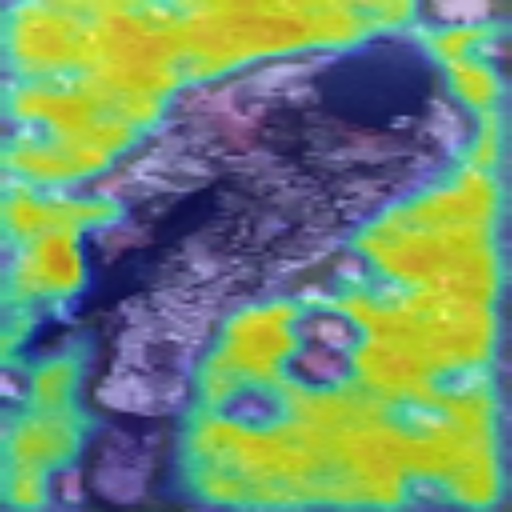} &
        \includegraphics[width=0.155\columnwidth]{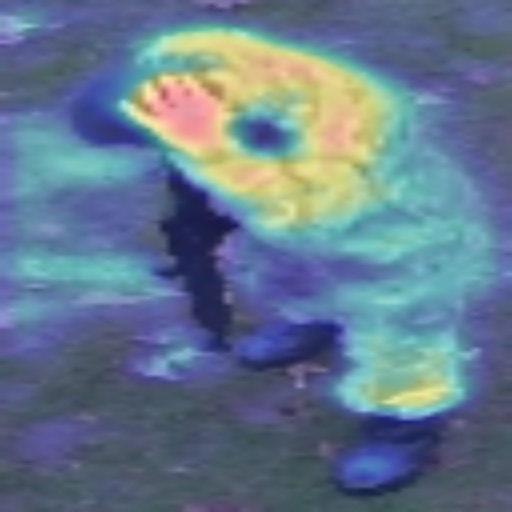} &
        \includegraphics[width=0.155\columnwidth]{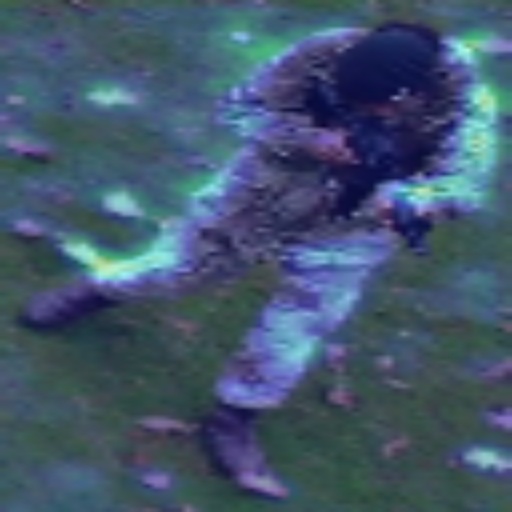} &
        \includegraphics[width=0.155\columnwidth]{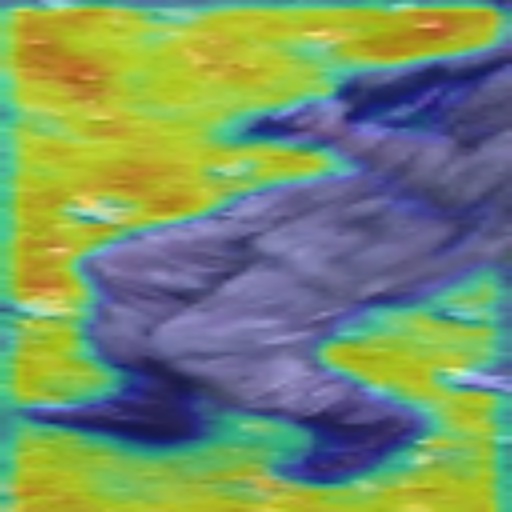} &
        \includegraphics[width=0.155\columnwidth]{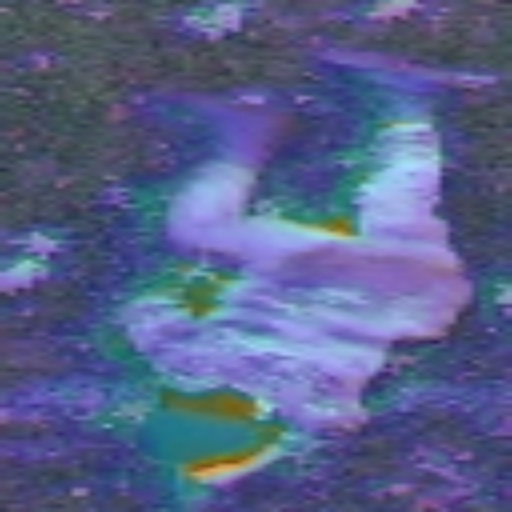} &
        \includegraphics[width=0.155\columnwidth]{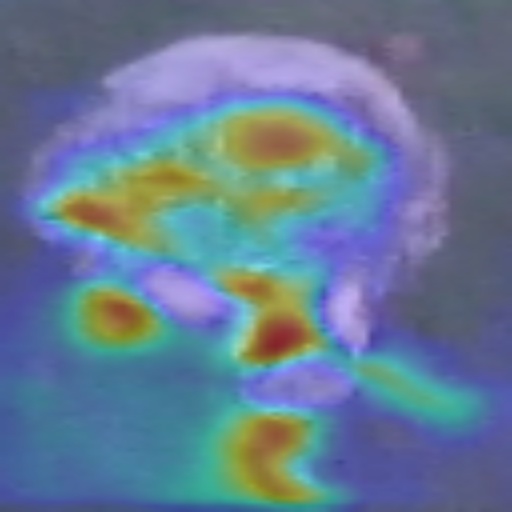} \\[-0.3mm]
        \multicolumn{6}{c}{\scriptsize (b) Ours 2-layer CNN} \\[0.6mm]

        % Row 3
        \includegraphics[width=0.155\columnwidth]{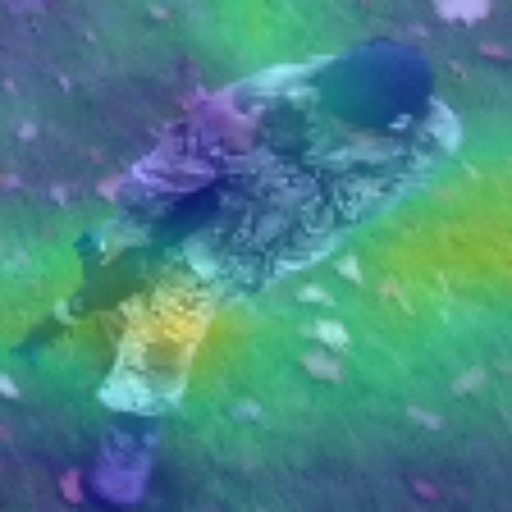} &
        \includegraphics[width=0.155\columnwidth]{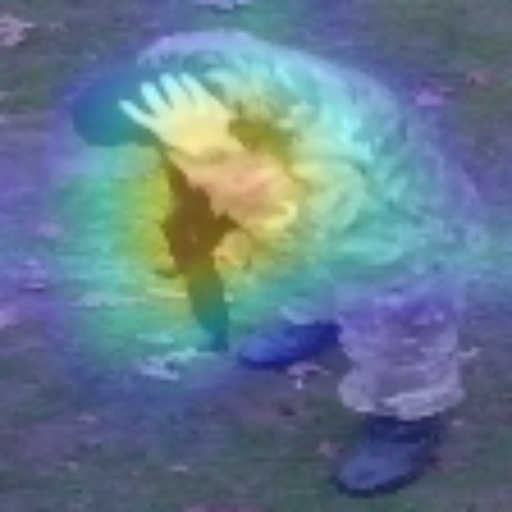} &
        \includegraphics[width=0.155\columnwidth]{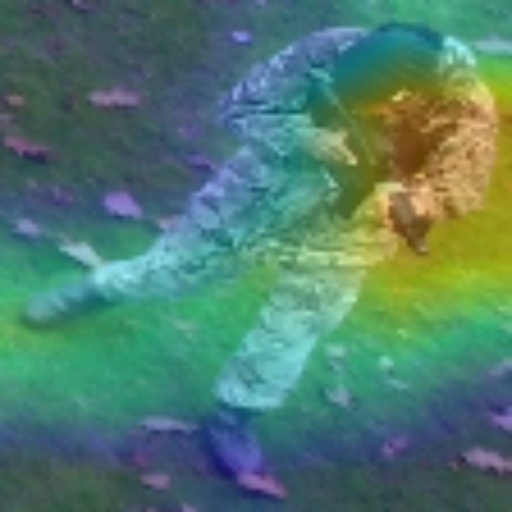} &
        \includegraphics[width=0.155\columnwidth]{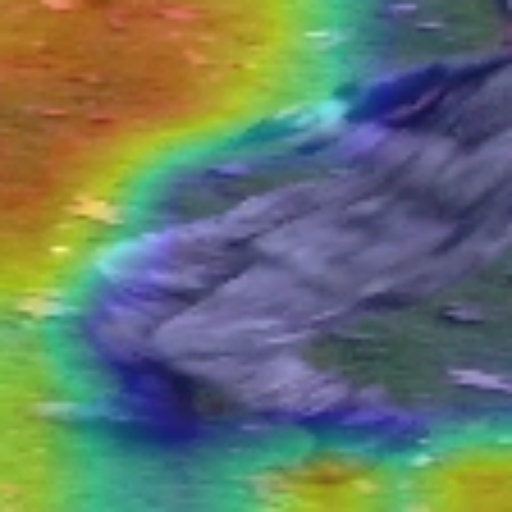} &
        \includegraphics[width=0.155\columnwidth]{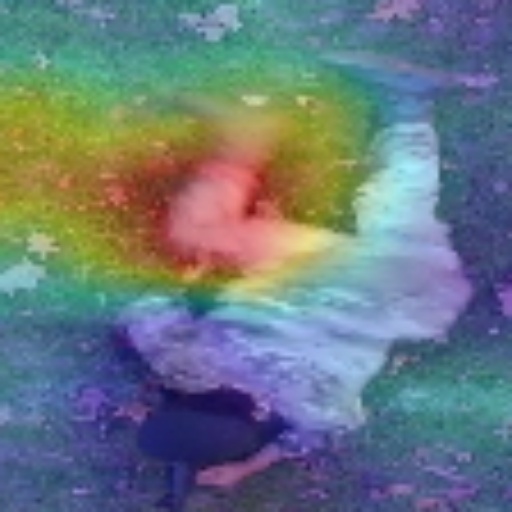} &
        \includegraphics[width=0.155\columnwidth]{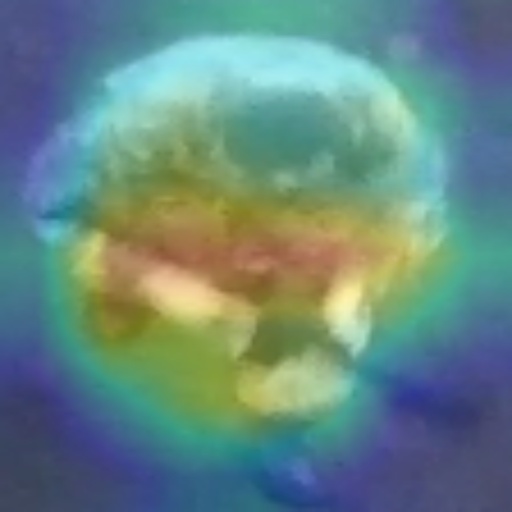} \\[-0.3mm]
        \multicolumn{6}{c}{\scriptsize (c) R(2+1)D}
    \end{tabular}%
    }
    \caption{Grad-CAM Visualisations for Explainability, showing that model learns features from the most relevant regions in the image.}
    \label{fig:gradcam}
\end{figure}

Grad-CAM (in Fig.~\ref{fig:gradcam}) provides important explainability of the ATRACT, as it reveals whether the model attends to behaviourally meaningful regions in the images rather than relying on spurious background patterns. In this sense, explainability supports the broader trustworthiness objective of ATRACT, where AI outputs should be interpretable to support human-in-the-loop remote casualty triage. For several actions, the proposed 2-layer CNN appears to attend more closely to fine-grained cues that are directly related to the performed actions. In case of head-injury, for instance, it places stronger emphasis on raised hand over the head, which is a more specific behavioural cue, whereas R(2+1)D tends to spread attention more broadly over the upper torso, arm, or wrist. Similar patterns are also visible in other classes. At the same time, visualisations also expose model weaknesses, such as arm-injury where CNN is partly influenced by background artifacts rather than focusing only on the subject. These observations are useful because they show not only where the models look, but also where attention diffuses or is less informative, a kind of evidence needed when assessing the robustness and trustworthiness of ATRACT system.

%%%%%%%%%%%%%%%%%%%%%%%%%%%%%%%%%%%%%%%%%%%%%%%%%%%5

% \vspace{-0.25cm}

\section{System Capacity and Complexity} \label{complexity}
For simplicity, we only highlight the computational complexity of pipeline during training and inference of the ATRACT system. Our system is relatively lightweight with fewer computational resources, requiring 17M parameters and 8.67 GFLOPs, compared with 32M parameters and 323 GFLOPs for R(2+1)D model. %This shows that proposed CNN-design requires substantially fewer computational resources, which is encouraging for a pilot triage system. 
At the same time, training and inference times are slightly higher in our setup (ours: 348s and 87s \emph{vs.} R(2+1)D: 248s and 73s), indicating that runtime is not determined by GFLOPs alone. Overall, these results suggest that ATRACT maintains a compact learning architecture while keeping manageable training and inference costs.

\begin{table}[!htbp]
\caption{Comparison of model complexity in terms of parameters, GFLOPs, train time, and inference time.}
\centering
\resizebox{\columnwidth}{!}{
\setlength{\tabcolsep}{14pt}
\renewcommand{\arraystretch}{1.35}
\begin{tabular}{lccccc}
\toprule
& \textbf{Param.} & \textbf{GFLOPs} & \textbf{Train Time} & \textbf{Infer. Time} \\
\midrule
Ours & 17M & 8.67 & 348s & 87s \\
R(2+1)D & 32M & 323 & 248s & 73s  \\
\bottomrule
\end{tabular}
\label{tab:complexity}
}
\end{table}

% \vspace{-0.2cm}

\section{Ethics}\label{sec:ethics}

Given the safety-critical nature of battlefield triage, ethical governance is a foundational feature of the ATRACT system. We set out to create a framework for trustworthiness in which technical robustness, legal compliance, and ethical accountability are maintained across design, development, testing, and prospective deployment. From the outset, ATRACT was intended to comply with the laws of armed conflict, medical ethics, and the UK Ministry of Defence AI Ethics Principles~\cite{taddeo2022ethical}: 1) Human-Centricity, 2) Responsibility, 3) Understanding, 4) Bias and Harm Mitigation, and 5) Reliability. These principles emphasise that AI-based systems should prioritise humanity, life, and wellbeing, while providing accountability through explicability, transparency, bias mitigation, operational robustness, and security.

Because multi-modal AI outputs may influence high-stakes triage decisions under conditions of uncertainty, particular emphasis is placed on data quality, bias mitigation, transparency, interpretability, reproducibility, and reliability. The system is therefore designed to support human decision-makers in a consistent and auditable manner, rather than operate as an opaque or fully autonomous agent. These commitments shaped data acquisition, model training, and system development. For example, written consent was obtained before drone filming of Combined Cadet Force participants, and flights were conducted by a qualified drone pilot under strict airspace and safety procedures. Since realistic injured-soldier body-sensor data are difficult and ethically constrained to collect, publicly available MIMIC data were used to augment physiological signals without involving wounded personnel. In parallel, transparent and reproducible multi-modal learning methods, alongside expert-in-the-loop evaluation, help ensure that ATRACT remains a decision-support framework in which ethical, legal, and technical safeguards are integrated throughout the research lifecycle. A full ethics checklist was created as part of the project and is included in the appendix of~\cite{lee2026ai}.

\section{Conclusion} \label{conclude}
This research presents \textit{ATRACT}, a multi-modal human-in-the-loop triage support system for remote casualty assessment in contested environments. By integrating drone-assisted visual sensing with wearable sensor input for multi-modal learning, the proposed system addresses the limitations of relying on a single source of evidence for casualty triage. We faced a unique challenge due to the absence of realistic injured body-sensor data (heart rate, breathing rate), and addressed the data realism gap by augmenting the recordings for injured actions. Experimental results on our drone captured dataset demonstrated that multi-modal fusion improves action recognition accuracy to \textbf{85.7}\%, while our lightweight visual encoder performed competitively with stronger pretrained backbones. Overall, findings support the feasibility and practicality of trustworthy multi-modal human-in-the-loop triage. \\
\textbf{Limitation:} Existing ATRACT remains limited for battlefield use cases at the moment, where robust remote assessment requires integrated visual, physiological, and human-in-the-loop decision support under uncertainty. Future work will investigate and incorporate other modalities (thermal) during a larger-scale data collection, improved generalisation across operational conditions, sophisticated human-in-the-loop supervision, and integration with the medics--decision support system.

\section*{Acknowledgments}
ATRACT was supported by UK Research and Innovation (UKRI) - Engineering and Physical Sciences Research Council (EPSRC) under Grant EP/X028631/1. For data collection, authors are thankful to cadets from Haberdashers’ Adams Grammer School, High Street, Newport, Shropshire.

% {\appendix[Proof of the Zonklar Equations]
% Use $\backslash${\tt{appendix}} if you have a single appendix:
% Do not use $\backslash${\tt{section}} anymore after $\backslash${\tt{appendix}}, only $\backslash${\tt{section*}}.
% If you have multiple appendixes use $\backslash${\tt{appendices}} then use $\backslash${\tt{section}} to start each appendix.
% You must declare a $\backslash${\tt{section}} before using any $\backslash${\tt{subsection}} or using $\backslash${\tt{label}} ($\backslash${\tt{appendices}} by itself
%  starts a section numbered zero.)}


% \section{References Section}
% You can use a bibliography generated by BibTeX as a .bbl file.
%  BibTeX documentation can be easily obtained at:
%  http://mirror.ctan.org/biblio/bibtex/contrib/doc/
%  The IEEEtran BibTeX style support page is:
%  http://www.michaelshell.org/tex/ieeetran/bibtex/
 
% You can manually copy in the resultant .bbl file and set second argument of $\backslash${\tt{begin}} to the number of references
%  (used to reserve space for the reference number labels box).

\bibliographystyle{IEEEtran}
\bibliography{refs}

\end{document}